\documentclass[ejs]{imsart}
\RequirePackage[OT1]{fontenc}
\RequirePackage{amsthm,amsmath}
\RequirePackage[numbers]{natbib}
\usepackage{graphicx,caption}
\RequirePackage[colorlinks,citecolor=blue,urlcolor=blue]{hyperref}
\usepackage{multirow,tabularx,booktabs,siunitx}
\usepackage{mathrsfs,amsmath,amsthm,amssymb,color}
\usepackage{enumitem}
\usepackage{bm}
\usepackage{algorithm}
\usepackage{algorithmic}

\startlocaldefs
\numberwithin{equation}{section}
\theoremstyle{plain}
\newtheorem{theorem}{Theorem}
\newtheorem{lemma}{Lemma}
\newtheorem{proposition}{Proposition}

\newtheorem{assumption}{Assumption}
\newtheorem{corollary}{Corollary}
\newtheorem{example}{Example}
\endlocaldefs

\def\beq{\begin{equation}}
\def\eeq{\end{equation}}
\def\beqr{\begin{eqnarray}}
\def\eeqr{\end{eqnarray}}
\def\beqrs{\begin{eqnarray*}}
\def\eeqrs{\end{eqnarray*}}
\def\bet{\begin{theorem}}
\def\eet{\end{theorem}}
\def\bel{\begin{lemma}}
\def\eel{\end{lemma}}
\def\bep{\begin{proposition}}
\def\eep{\end{proposition}}
\def\bep{\begin{assumption}}
\def\eep{\end{assumption}}

\def\wt{\widetilde}

\def\wh{\widehat}

\def\one{{\bf 1}}

\def\mB{\mathbb{B}}
\def\mC{\mathcal C}

\def\mE{\mathbb{E}}

\def\mI{\mathcal I}
\def\bI{\mathbb I}
\def\mJ{{\mathcal{J}}}
\def\mM{\mathcal M}

\def\mR{\mathbb{R}}
\def\mS{\mathcal S}

\def\argmax{\mbox{argmax}}

\setlist[enumerate]{label*=(A\arabic*)} 

\begin{document}

\begin{frontmatter}
\title{Subsampling-Based Modified Bayesian Information Criterion for Large-Scale Stochastic Block Models
\support{Danyang Huang's research is partially supported by the National Natural Science Foundation of China (grant numbers 12071477,
11701560), as well as the Fund for building world-class universities (disciplines) of Renmin University of China. Bo Zhang's research is partially supported by the National Natural Science Foundation of China (grant number, 71873137), as well as the Fund for building world-class universities (disciplines) of Renmin University of China. The authors gratefully acknowledge the support of Public Computing Cloud, Renmin University of China.}}
\runtitle{Subsampling-Based Modified Bayesian Information Criterion}
\runauthor{Deng et al.}

\begin{aug}
\author{\fnms{Jiayi} \snm{Deng}}
\address{Center for Applied Statistics and School of Statistics, Renmin University of China}

\author{\fnms{Danyang} \snm{Huang}
\ead[label=e1]{dyhuang@ruc.edu.cn}}
\address{Center for Applied Statistics and School of Statistics, Renmin University of China
\printead{e1}}

\author{\fnms{Xiangyu} \snm{Chang}}
\address{School of Management, Xi'an Jiaotong University}

\author{\fnms{Bo} \snm{Zhang}
\ead[label=e2]{mabzhang@ruc.edu.cn}}
\address{Center for Applied Statistics and School of Statistics, Renmin University of China \printead{e2}}

\end{aug}

\begin{abstract}
Identifying the number of communities is a fundamental problem in community detection, which has received increasing attention recently. However, rapid advances in technology have led to the emergence of large-scale networks in various disciplines, thereby making existing methods computationally infeasible. To address this challenge, we propose a novel subsampling-based modified Bayesian information criterion (SM-BIC) for identifying the number of communities in a network generated via the stochastic block model and degree-corrected stochastic block model. We first propose a node-pair subsampling method to extract an informative subnetwork from the entire network, and then we derive a purely data-driven criterion to identify the number of communities for the subnetwork. In this way, the SM-BIC can identify the number of communities based on the subsampled network instead of the entire dataset. This leads to important computational advantages over existing methods. We theoretically investigate the computational complexity and identification consistency of the SM-BIC. Furthermore, the advantages of the SM-BIC are demonstrated by extensive numerical studies.
\end{abstract}

\tableofcontents
\end{frontmatter}

\section{Introduction}

Network community detection is one of the most widely-studied topics in network analysis \citep{girvan2002community, newman2004finding, fortunato2010community}. Intuitively, for networks with assortative communities, community detection aims to distribute the network nodes to several clusters, so that nodes in the same cluster have denser connectivity. Network community structure is beneficial for understanding the characteristics of each cluster \citep{girvan2002community, bickel2009nonparametric}. Specifically, in social network platforms (e.g., {\it Facebook}, {\it Twitter}, and {\it Sina Weibo}), communities can be formed by users with similar interests or preferences, which enables online platforms to recommend suitable products and services to targeted groups \citep{bodapati2008recommendation, homburg2021wage, shaddy2022express}.

In the past few decades, numerous assortative community detection methods have been proposed, including but not limited to modularity maximization \citep{newman2006modularity, good2010performance}, spectral clustering \citep{ng2002spectral, von2007tutorial, rohe2011spectral}, belief propagation \citep{hastings2006community, yedidia2003understanding}, and pseudo-likelihood methods \citep{amini2013pseudo, wang2021fast}. Theoretically, the stochastic block model (SBM), has been widely assumed to analyze the consistency properties of network community methods \citep{holland1983stochastic, snijders1997estimation, nowicki2001estimation}. It should be noted that most community detection methods require the number of communities $K_0$ to be known in advance. Then, the theoretical properties can be carefully established. However, $K_0$ is typically unknown in real-world networks. Therefore, how to choose $K_0$ is important.

A variety of methods have been proposed to determine the number of communities $K_0$, such as the eigenvalue-based methods \citep{le2015estimating, bordenave2015non, bickel2016hypothesis}, semi-definite programming-based methods \citep{li2013revealing, yan2018provable}, network cross-validation methods \citep{chen2018network, li2020network}, and likelihood-based methods \citep{daudin2008mixture, wang2017likelihood, hu2020corrected, ma2021determining}. Specifically, the eigenvalue-based methods estimate the number of communities based on the eigenvalue properties of non-backtracking, Bethe Hessian, or normalized Laplacian matrices \citep{le2015estimating, bordenave2015non, bickel2016hypothesis, dall2021unified}. Additionally, the semi-definite programming approach identifies $K_0$ by solving a semi-definite optimization problem \citep{li2013revealing, yan2018provable}. Moreover, the network cross-validation method extends the cross-validation method to network data via a network sampling strategy \citep{chen2018network, li2020network}. Lastly, the likelihood-based approaches aim to make full use of observed samples, which have been widely studied, including Bayesian information criterion and likelihood ratio methods. Specifically, the Bayesian information criterion consists of a conditional log-likelihood of entire observations and a penalty term that depends on the prior distribution of the latent variable \citep{daudin2008mixture, saldana2017many, hu2020corrected}. For the likelihood ratio approaches, \cite{ma2021determining} proposed to estimate $K_0$ by comparing the goodness-of-fit of two models estimated by a candidate number of communities $K$ and $K+1$. Moreover, \cite{wang2017likelihood} discussed the asymptotic properties of the log-likelihood ratio statistics.
It is remarkable that to evaluate each candidate $K$ via the aforementioned criteria, such as the network cross-validation methods and the likelihood-based approaches, we need to first estimate the parameters for the SBM using the entire observed network.
In this case, spectral clustering is considered a simple and easy-to-implement approach with well-founded theoretical guarantees \citep{rohe2011spectral, chaudhuri2012spectral, zhao2012consistency, lei2015consistency}.

However, recent advances in science and technology have brought about large-scale network data, leading to unprecedented computational challenges for community detection. For example, as reported by {\it Statista} ({\it www.statista.com}), in January 2022, the online social networks {\it Facebook}, {\it Twitter}, and {\it Sina Weibo} had approximately 2,910 million, 436 million, and 573 million active users, respectively. Researchers could also access the relationships of millions of network nodes using open-source datasets, such as the {\it Stanford Large Network Dataset}\footnote{\url{http://snap.stanford.edu}}, which has collected different networks with more than 10 million nodes each. Consequently, directly applying traditional methods to estimate $K_0$ for these large-scale network data is impractical.
For example, for a network with $N$ nodes, the time complexity of spectral clustering-based methods is no lower than $O(N^{3})$ for estimating $K_0$ \citep{yan2009fast, li2011time, chen2011large}. Even if the algorithm could be accelerated, the computational complexity is still in the order of $O(N^{2})$ \citep{halko2011finding, feng2018faster, martin2018fast}. To deal with the computational challenge brought by large-scale networks, subsampling is a valuable tool \citep{politis1999subsampling}. Its main advantage is that we can obtain a computationally efficient and consistent estimator based on a small subsample \citep{wang2018optimal, wang2019information, wang2021optimal, yu2022optimal}. Although subsampling pays the price of statistical convergence, it makes the traditional methods feasible in large-scale data analysis.

In the literature, various sampling designs have been proposed to derive representative samples of a given network, which include node sampling methods \citep{snijders1999non, bhattacharyya2015subsampling, mukherjee2021two} and edge sampling methods \citep{goldreich2008approximating, gonen2011counting, li2020network}. The node sampling methods select landmark nodes from the entire network, and the subnetwork is induced by these selected nodes. Uniform node sampling is considered to be the simplest method and has been widely used \citep{snijders1999non, bhattacharyya2015subsampling, lunde2019subsampling,mukherjee2021two}. Another widely studied node sampling method is snowball sampling \citep{illenberger2012estimating, pattison2013conditional, chen2019bootstrap}. Based on the snowball sampling approach, \cite{thompson2016using} and \cite{akcora2019graphboot} recently developed bootstrap methods to reduce estimation bias for large networks. The edge sampling methods randomly collect edge samples from the entire network, which have also received considerable attention \citep{feige2004sums, eden2017approximately, li2020network}. For example, \cite{feige2004sums} and \cite{goldreich2008approximating} adopted edge sampling procedures in estimating the average degree of a network. Recently, edge sampling approaches have been investigated to approximate counting the number of subgraphs \citep{gonen2011counting,eden2017approximately,assadi2018simple}. Moreover, \cite{li2020network} applied uniform edge sampling in random graph model selection. Note that existing studies focus on subsampling many times to provide stable statistical inference for network models. However, we aim to conduct subsampling only once to allow model selection for large-scale networks with limited computational resources.

This work proposes a novel subsampling-based modified Bayesian information criterion (SM-BIC) for identifying the number of communities for large-scale SBMs. Specifically, in the context of large-scale networks, we first develop a {\it node-pair subsampling} method to extract a subnetwork from the entire network. The node-pair subsampling method combines the idea of uniform node sampling and edge sampling. More precisely, we first uniformly and randomly select a subset of nodes from the entire network and then collect all edges related to these nodes to construct a subnetwork. In this way, this subnetwork fully retains the connection information between the selected nodes and the entire network. Note that the node-pair subsampling method only requires subsampling once due to computational efficiency. Then, based on the selected subnetwork, we derive a purely data-driven criterion without tuning any parameters.
Since the criterion is based only on subsampled data, it makes the subsequent parameter estimation applicable even for large-scale networks with affordable computational resources. In particular, we use spectral clustering for the subsampled subnetwork to obtain the community assignments. In this way, the computational complexity of the SM-BIC can be as low as $O(Nn)$, where $n$ is the subsample size satisfying $n<<N$. Furthermore, we extend the SM-BIC to the degree-corrected stochastic block model (DCSBM) \citep{karrer2011stochastic}. We theoretically investigate the computational advantage of the SM-BIC. Most importantly, for both the SBM and DCSBM, we establish the consistency of the SM-BIC by studying the penalized log-likelihood function under misspecification cases (e.g., under-fitting and over-fitting).

To summarize, the proposed method has the following advantages. First, compared with the eigenvalue-based methods \citep{bordenave2015non, le2015estimating, bickel2016hypothesis, dall2021unified}, the SM-BIC fully exploits the connectivity information in the selected subnetwork, while the eigenvalue-based methods use the eigenvalue information of network matrices. Second, compared with the method based on semi-definite programming \citep{li2013revealing, yan2018provable}, the proposed SM-BIC method applies the spectral clustering algorithm to identify community labels for network nodes, which is more computationally efficient. Third, compared with the network cross-validation methods \citep{chen2018network, li2020network}, the SM-BIC only requires subsampling once, while the network cross-validation method uses a network resampling technique, which requires tuning the number of folds. Finally, compared with the aforementioned BIC-based approaches \citep{daudin2008mixture, saldana2017many, hu2020corrected} and likelihood ratio methods \citep{wang2017likelihood, ma2021determining}, the SM-BIC can identify $K_0$ using only a small subnetwork; further, it is a completely data-driven method without any predefined tuning parameters. Consequently, the SM-BIC could be feasibly applied to identify the number of communities for large-scale networks with affordable computational resources. Specifically, its computational complexity could be as low as $O\{N(\log{N})^2\}$, as demonstrated in Propositions \ref{pro: computational} and \ref{pro: size}.

The remainder of this paper is organized as follows. In Section \ref{sec: subsampling}, we introduce the subsampling-based modified Bayesian information criterion. In Section \ref{sec: theoretical}, we discuss the theoretical properties of the SM-BIC and establish the consistency of the estimator of the number of communities. In Section \ref{sec: numerical}, we demonstrate the effectiveness of our method through extensive numerical studies. Further discussions are provided in Section \ref{sec: concluding}. Proofs are presented in the Appendices and the supplementary materials.

\section{Subsampling-based modified Bayesian information criterion for stochastic block model}\label{sec: subsampling}

In this section, we first introduce the stochastic block model and challenges of existing model selection methods. Then, we develop the SM-BIC for large-scale SBMs and extend the criterion to DCSBMs. Lastly, we discuss the parameter estimation procedure for this method.

\subsection{Preliminaries}

Consider a large-scale undirected graph generated from an SBM with $N$ nodes and $K_0$ communities. The observed random graph is often represented by a symmetric adjacency matrix $A \in \mR^{N\times N}$ with zero diagonal entries. Specifically, for any node pair $(i,j)$, if there is a connection, then $A_{ij}=1$; otherwise, $A_{ij}=0$. For each node $i$, denote its community label as $g^{*}_{N,i} \in [K_0]=\{1,\cdots, K_0\}$. Let $N_{k,g^{*}_N}= \sum_{i} \bI(g^{*}_{N,i}=k)$ denote the size of the $k$-th cluster. Given a label vector $g^{*}_{N}= (g^{*}_{N,1}, \cdots, g^{*}_{N,N})^\top\in [K_0]^N$, the edge variables $A_{ij}$s for $i<j$ are independent Bernoulli random variables with $\mE(A_{ij}) = B^{*}_{g^{*}_{N,i}g^{*}_{N,j}},$ where $B^{*}=(B^{*}_{kl})\in (0,1)^{K_0\times K_0}$ is a symmetric matrix describing connectivity probability within and between communities. Namely, each element $B^{*}_{kl}\in(0,1)$ represents the connectivity probability between $k$ and $l$ communities ($1\leq k,l\leq K_0$). In this way, the connectivity probability between any node pair ($i, j$) depends only on their community labels. For simplicity, let ${\rm SBM}_{K_0}(g^{*}_{N}, B^{*})$ represent a stochastic block model with $K_0$ blocks parameterized by $g^{*}_{N}$ and $B^{*}$.

Throughout this paper, we let $g^{*}_{N}$ and $B^{*}$ denote the true parameters of the observed adjacency matrix $A$. Furthermore, $K_0$ is considered to be a fixed constant. For any $1\le l \neq k \le K_0$, we assume $B^{*}_{kl} < B^{*}_{kk}$, which means that the connectivity probability of within-community is higher than that of between-community. Under any candidate $K$, denote $g_{N} \in [K]^{N}$ as the community assignment of the $K$-block model, and the corresponding connectivity matrix is represented by a symmetric matrix $B\in \mB_{K}= (0,1)^{K\times K}$. Additionally, when we refer to model selection, we mean the selection of $K_0$ for ${\rm SBM}_{K_0}(g^{*}_{N}, B^{*})$.

For the likelihood-based methods, to determine the number of communities, it is necessary to estimate the community assignment $g_N$ for each candidate $K$. For super-large $N$, even if accelerated algorithms are adopted, the computational cost is still high. For example, the randomized spectral clustering algorithm \citep{zhang2020randomized} has computational cost in the order of $O(N^{2})$. This motivates us to develop a network subsampling-based model selection criterion that reduces the cost by investigating small subsamples.

\subsection{Subsampling-based modified Bayesian information criterion}

\begin{figure}
 \centering
 \includegraphics[width=0.65\linewidth]{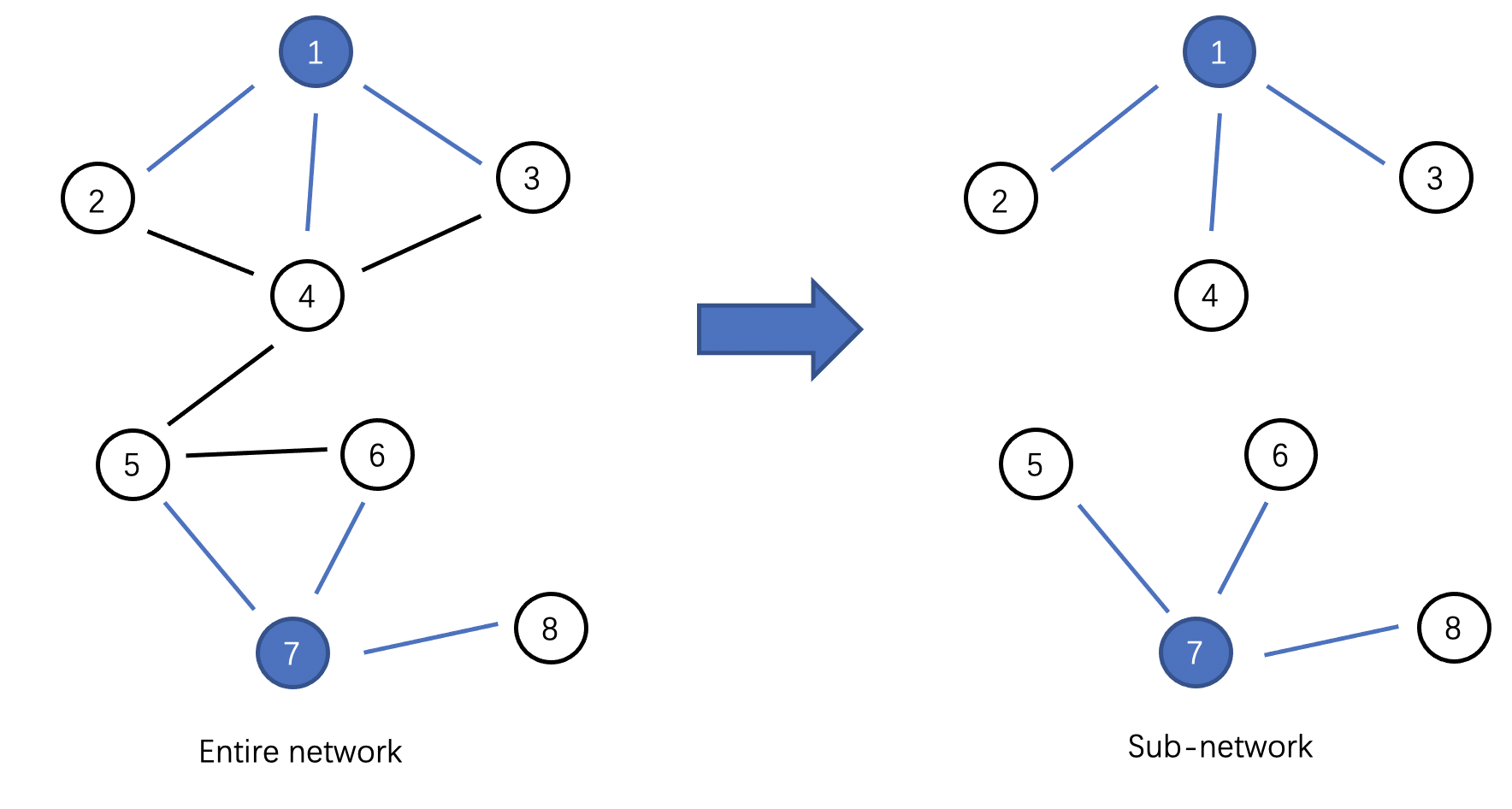}
 \caption{An example of {\it node-pair subsampling}. The left panel shows the entire network, where the colored nodes are considered to be selected by simple random sampling, whereas their corresponding connections (shown in dark blue) are extracted from the entire network. The right panel represents the subnetwork obtained by the {\it node-pair subsampling} method.}\label{fig: subsampling}
\end{figure}

In the context of large-scale networks, we first introduce the network subsampling method. Note that, unlike independent data, network data are correlated with each other by connections. To characterize the community membership of network nodes, we use a node-pair subsampling method to collect a subnetwork from the entire network. Specifically, we first uniformly sample $n$ nodes from $[N]$; that is, the probability of each node being selected is equal to $n/N$, where the subsample size $n <<N$. We further denote the set of selected nodes as $\mS=\{j\in [N]: \text{node} \ j \ \text{is} \ \text{selected}\}$. Then, we sample all node pairs related to these selected nodes. That is, if node $i$ is selected and there is a connection between $i$ and $j$, then node pair $(i, j)$ is also collected. The subsampling method is illustrated in Figure \ref{fig: subsampling}. We refer to this method as {\it node-pair subsampling}. For convenience, let $s_j$ ($s_j\in [n]$) denote the index of the selected node $j$ in the node set $\mS$. Define a $N\times n$ matrix $A^{\mS}$ to represent these selected connections, where the entries are $A^{\mS}_{is_j}=A_{ij}$, for $i \in [N], j\in \mS$. Then, we focus on the observation $A^{\mS}$ rather than the entire network connections, to identify the number of communities.

For model selection, we introduce the proposed modified Bayesian information criterion based on $A^{\mS}$. The criterion is derived from the maximization of the log-posterior likelihood function of $g_N$. We first provide the prior distribution of $g_N$ under ${\rm SBM}_{K}$. Based on the selected sample $A^{\mS}$, we demonstrate that the community partition of the entire network is determined by the community assignment of the selected nodes. Specifically, consider the community assignment of the selected nodes to be $g_{n}$ ($g_{n}\in [K]^{n}$), and $g_{n,s_j}$ is the community label of the selected node $j$. Then, for any unselected node $i\notin \mS$, we have different ways to obtain its community label based on the label of the selected nodes. For example, we could cluster this node to the community with the most connections to it. Namely, the community label of the unselected node $i$ is given by $\wh{g}_{N,i}=\max_{k}\sum_{j\in \mS}A_{ij}\bI(g_{n,s_j}=k)$, where $\bI(\cdot)$ is an indicator function. We could alternatively obtain the label assignment for unselected nodes by spectral clustering, which is illustrated in detail in the next subsection. In this way, based on $A^{\mS}$, the set of all possible community assignments for entire network nodes is provided as $\mC(A^{\mS},K)= \bigcup_{g_{n}\in [K]^{n}}\big\{g_N \in [K]^{N}: \forall \ i \notin \mS, g_{N,i}=\max_{k}\sum_{j\in \mS}A_{ij}\bI(g_{n,s_j}=k), \forall \ j \in \mS, g_{N,j}= g_{n,s_j}\big\}$. Therefore, the number of possible community assignments is $|\mC(A^{\mS},K)|= K^{n}$. Similar to \cite{chen2008extended}, we assign the prior probability to $g_N$ as
 \beq\label{eq: prior}
 \phi(g_N)=K^{-n},\ \text{for} \ g_N \in \mC(A^{\mS},K).
 \eeq
 Next, we analyze the posterior probability of $g_N$.

We start with studying the probability of $A^{\mS}$ under ${\rm SBM}_{K}$. We denote the set of node pairs corresponding to the independent edge variables in $A^{\mS}$ by $E=E_{\rm in}\cup E_{\rm out}$. Where $E_{\rm in}=\{(i,j): i,j \in \mS, j>i\}$ and $E_{\rm out}=\{(i,j): i \in [N]-\mS, j \in \mS\}$ represent the set of node pairs within selected nodes and that between selected and unselected nodes, respectively. Moreover, since $|E_{\rm in}|=n(n-1)/2$ and $|E_{\rm out}|=(N-n)n$, we have $|E|=Nn- n(n+1)/2$. Let $o_{kl,g_N}=\sum_{(i,j)\in E} A_{ij}\bI(g_{N,i}=k, g_{N,j}=l)$ and $n_{kl,g_N}= \sum_{(i,j)\in E}\bI(g_{N,i}=k, g_{N,j}=l)$ denote the number of observed connections and the number of maximum possible connections between $(k,l)$ clusters, respectively. Additionally, define a vector $\theta \in \Theta_{K}=(0,1)^{K(K+1)/2}$ to represent the upper triangle elements of $B$. Then, given ($g_N, \theta$), the log-likelihood function of $A^{\mS}$ is $$\log{f(A^{\mS}|g_N, \theta)}= \sum_{1\le k\le l\le K} \{ o_{kl,g_N}\log{ \theta_{kl}} + (n_{kl,g_N}- o_{kl,g_N})\log{(1-\theta_{kl})} \}.$$ Accordingly, the likelihood function of $g_N$ is  $f(A^{\mS}|g_N) = \int f(A^{\mS}|g_N,\theta)p(\theta)\mbox{d}\theta$, where $p(\theta)$ is the prior distribution of $\theta$.

Then, we give an approximation of the log-likelihood function $\log{f(A^{\mS}|g_N)}$ in the following lemma.
\begin{lemma}[\textbf{Log-likelihood function approximation}]\label{lem: log-likelihood} Suppose the adjacency matrix $A$ generated from ${\rm SBM}_{K}$ and the subset of nodes $\mS$ collected by simple random sampling $n$ nodes from the entire network. Then, the log-likelihood function $\log{f(A^{\mS}|g_N)}$ can be approximated by,
\beq\label{eq: likelihood}
\log{f(A^{\mS}|g_N)}= \sup_{\theta \in \Theta_K} \log{f(A^{\mS}|g_N,\theta)} - \frac{K(K+1)}{4} \log{M}+O(1),
\eeq
where $M$ denotes the number of independent edge variables in $A^{\mS}$, i.e., $M=|E|=Nn- n(n+1)/2$.
\end{lemma}
\noindent The proof of Lemma \ref{lem: log-likelihood} can be found in Appendix \ref{appb: likelihood}. As a result, under ${\rm SBM}_{K}$, according to \eqref{eq: prior} and \eqref{eq: likelihood}, the log-posterior probability of $g_N$ is
\beq\label{eq: posterior}
\log{f(g_N| A^{\mS})}=\log\{f(A^{\mS}|g_N)\phi(g_N)\} + c,
\eeq
where $c=-\int \log\{f(A^{\mS}|g_N)\phi(g_N)\} \mbox{d}g_N$ is a constant.

We now establish the SM-BIC. According to Bayesian inference, the community assignment that maximizes the posterior probability is estimated, that is $\wh{g}_N= \argmax_{g_N\in \mC(A^{\mS},K)} \log{f(g_N|A^{\mS})}.$ To this end, based on \eqref{eq: likelihood} and \eqref{eq: posterior}, the SM-BIC is proposed as follows:
\beq\label{eq: smbic}
\ell(K)= \max_{g_N\in \mC(A^{\mS},K)} \sup_{B\in \mB_{K}}\log{f(A^{\mS}|g_N,B)} -\left\{n\log{K}+ \frac{K(K+1)}{4}\log{M}\right\}.
\eeq
\noindent The form of the criterion \eqref{eq: smbic} seems to be similar to the corrected BIC criterion proposed by \citep{hu2020corrected}. However, there are two key differences from the corrected BIC, which are also the key contributions of our criterion. First, the SM-BIC is a purely data-driven method without any predefined tuning parameters, whereas the corrected BIC requires choosing one parameter to control the model selection results. This is because we assume a simple uniform prior for ${\rm SBM}_{K}$ and the latent label vector $g_N$; this prior setting follows the work of \cite{chen2008extended}.
Second, based on \eqref{eq: smbic}, we estimate the community assignment from $A^{\mS}$, which has a lower dimension than $A$ for $n<<N$. Hence, criterion \eqref{eq: smbic} could save computational costs. We demonstrate the important computational advantages of the SM-BIC in Subsection \ref{subsec: parameter}.

\subsection{Extension to degree-corrected stochastic block model}

The DCSBM \citep{karrer2011stochastic} is generalized from the SBM, which introduces node-specific parameters to allow for degree heterogeneity within communities. Specifically, given parameters $g_N, B$, the probability of an edge between $(i, j)$ is represented by $ P(A_{ij}=1)= \psi_i B_{g_{N,i}g_{N,j}} \psi_j$, where the parameter $\psi_i$ characterizes the individual activeness of node $i$. In this way, a DCSBM is parameterized by a triplet $(g_N, B, \psi)$ where $\psi=(\psi_1,\cdots, \psi_N)^{\top}$. For consistency, we assume that the underlying model is ${\rm DCSBM}_{K_0}(g^{*}_N,B^{*},\psi^{*})$. For identifiability of this model, the constraint $\sum_{i}\psi_i^{*} \bI(g^{*}_{N,i}=k)=N_{k,g^{*}_N}$ is imposed on each community $1\le k \le K_0$. Then, we extend the SM-BIC to the DCSBM.

We start with the log-likelihood function of the subsampled adjacency matrix $A^{\mS}$. Similar to \cite{karrer2011stochastic} and \cite{zhao2012consistency}, we replace Bernoulli likelihood with Poisson likelihood and assume
$A_{ij} \sim {\rm Poisson}( \psi_i B_{g_{N,i}g_{N,j}} \psi_j)$ to simplify the derivation. Furthermore, let $n_{kl,g_N}(\psi)= \sum_{(i,j)\in E}\psi_i\psi_j\bI(g_{N,i}=k, g_{N,j}=l)$. In this way, under ${\rm DCSBM}_{K}(g_N,B,\psi)$, the log-likelihood function of the subsampled adjacency matrix $A^{\mS}$ is given by $$ \log{f(A^{\mS} |g_N,B,\psi)}= \sum_{(i,j)\in E}A_{ij} \log{(\psi_i\psi_j)}+ \sum_{1\le k\le l\le K}\{o_{kl,g_N}\log{B_{kl}}- n_{kl,g_N}(\psi)B_{kl}\}.$$

Then, we consider $\psi$ in two cases. First, if $\psi$ is known, according to \eqref{eq: smbic}, the SM-BIC of the DCSBM is proposed as follows:
\beq\label{eq: dc-smbic}
\ell(K)= \max_{g_N\in \mC(A^{\mS},K) }\sup_{B\in \mB_K} \log{f(A^{\mS}| g_N, B, \psi)}-\left\{n\log{K}+ \frac{K(K+1)}{4}\log{M}\right\}.
\eeq
 Second, if $\psi$ is unknown, we take a plug-in estimator $\wh{\psi}$ into the \eqref{eq: dc-smbic} criterion to replace $\psi$. In this case, an estimation of $\psi$ is provided in the following subsection.

\subsection{Parameter estimation based on subsampled adjacency matrix}\label{subsec: parameter}

Here, we first introduce how to apply the SM-BIC to determine the number of communities for large-scale SBMs. Specifically, based on {\it node-pair} subsampling, we evaluate each candidate $K$ through the following three steps: label assignment, parameter estimation, and SM-BIC calculation. Thereafter, we further present the estimation method of the degree heterogeneity cases.

\textbf{ Label assignment.} We first perform the label assignment step on the $N\times n$ subsampled adjacency matrix. For a candidate $K$ and subsampled adjacency matrix $A^{\mS}$, the extended spectral clustering algorithm can be accomplished as follows.
\begin{itemize}
 \item[(1)] Perform SVD on $A^{\mS}$, and extract the largest $K$ left eigenvectors, denoted as $V_1,\cdots, V_{K}$, and define a $N \times K$ matrix $V=(V_1,\cdots, V_{K})$ to represent the embedding matrix.
 \item[(2)] Apply K-means clustering to the rows of $V$ to estimate node assignments and denote the clustering results by $\wh{g}_N$.
 \end{itemize}

\textbf{ Parameter estimation}. Based on the estimated label vector $\wh{g}_N$, we construct the plug-in estimator for the connectivity matrix $B$. Specifically, for all $1\le k \le l \le K$, the estimated $(k, l)$-th element of $\wh{B}$ is
\beq\label{eq: bEstimator}
\wh{B}_{kl}= \frac{o_{kl,\wh{g}_N}}{n_{kl,\wh{g}_N}}=\frac{\sum_{(i,j)\in E} A_{ij}\bI(\wh{g}_{N,i}=k,\wh{g}_{N,j}=l)}{ \sum_{(i,j)\in E}\bI(\wh{g}_{N,i}=k,\wh{g}_{N,j}=l)},
\eeq
and taking $\wh{B}_{lk}=\wh{B}_{kl}$, we obtain the estimated connectivity matrix $\wh{B}$.

\textbf{SM-BIC calculation}. Given $(\wh{g}_N, \wh{B})$, we evaluate the estimated ${\rm SBM}_{K}(\wh{g}_N,\wh{B})$ by
\beq\label{eq: appsbic}
\wh{\ell}(K)= \log{f(A^{\mS}| \wh{g}_N, \wh{B})} - \left\{ n\log{K}+ \frac{K(K+1)}{4}\log{M}\right\}.
\eeq
Therefore, we choose $K$ which maximizes the SM-BIC \eqref{eq: appsbic} as the number of communities.

\begin{algorithm}[h]
\caption{Model Selection Algorithm for SBM}
\begin{algorithmic}
\STATE \textbf{Input}: adjacency matrix $A^{\mS}$, a maximum candidate $K_{\rm max}$.
\begin{itemize}
\item[1.] For each candidate $1\le K \le K_{\rm max}$,
\begin{itemize}
\item[1.1] (\textbf{Label Assignment}) compute the community assignment estimator $\wh{g}_N$ using spectral clustering on $A^{\mS}$;
\item[1.2] (\textbf{Parameters Estimation}) calculate the plug-in estimator $\wh{B}$ defined in \eqref{eq: bEstimator};
\item[1.3] (\textbf{SM-BIC Calculation}) calculate the SM-BIC $\wh{\ell}(K)$, defined in \eqref{eq: appsbic}.
\end{itemize}
\item[2.] Calculate $\wh{K}= \argmax_{1\le K \le K_{\rm max}}\wh{\ell}(K)$.
\end{itemize}
\STATE \textbf{Output}: the optimal choice of the number of communities, $\wh{K}$.
\end{algorithmic}\label{alg: sbm}
\end{algorithm}

In the framework of the DCSBM, we need to modify the parameters' estimation methods. First, under candidate ${\rm DCSBM}_{K}$, to obtain $\wh{g}_N$, we use the spherical spectral clustering method proposed by \citep{lei2015consistency}. Specifically, let $v_i$ be the $i$-th row of $V$, i.e., $V=(v_1, \cdots, v_N)^{\top}.$ Furthermore, let $\wt{V}$ be the row-normalized version of $V$, namely, the $i$-th row of $\wt{V}$ is $v_{i}/\|v_{i}\|$, where $\|\cdot \|$ denotes the Euclidean norm of a vector. Then, we estimate the node assignments by the following steps: (1) form matrix $\wt{V}$ by normalizing each row of $V$ to unit norm; and (2) perform K-means clustering to the rows of $\wt{V}$ to obtain $\wh{g}_N$.
Second, based on the embedding matrix $V$, the plug-in estimator of $\psi_i$ is provided as $\wh{\psi}_i= \|v_i\|$. Third, for $1\le k \le l \le K$, the estimated $(k,l)$-th entry of $B$ is given by,
\beq\label{eq: mbEstimator}
\wh{B}_{kl}=\frac{o_{kl,\wh{g}_N}}{n_{kl,\wh{g}_N}(\wh{\psi})}=\frac{\sum_{(i,j)\in E} A_{ij}\bI(\wh{g}_{N,i}=k,\wh{g}_{N,j}=l)}{ \sum_{(i,j)\in E} \wh{\psi}_i \wh{\psi}_{j}\bI(\wh{g}_{N,i}=k,\wh{g}_{N,j}=l)},
\eeq
and then take $\wh{B}_{lk}=\wh{B}_{kl}$. To this end, we obtain the SM-BIC for ${\rm DCSBM}_{K}$ by taking $(\wh{g}_N, \wh{B}, \wh{\psi})$ into \eqref{eq: dc-smbic}.

\begin{algorithm}[h]
\caption{ Model Selection Algorithm for DCSBM}
\begin{algorithmic}
\STATE \textbf{Input}: adjacency matrix $A^{\mS}$, a maximum candidate $K_{\rm max}$.
\begin{itemize}
\item[1.] For each candidate $1\le K \le K_{\rm max}$,
\begin{itemize}
\item[1.1] (\textbf{Label Assignment}) compute the membership labels estimator $\wh{g}_N$ by performing spherical spectral clustering on $A^{\mS}$;
\item [1.2] (\textbf{Parameter Estimation}) obtain $\wh{\psi}$ and $\wh{B}$ by the following steps,
\begin{itemize}
\item[(a)] compute the Euclidean norm of each row of matrix $V$, and then $\wh{\psi}_i=\|v_i\|$ for all $1\le i \le N$;
\item[(b)] calculate the plug-in estimator defined in \eqref{eq: mbEstimator};
\end{itemize}
\item[1.3] (\textbf{SM-BIC Calculation}) calculate the SM-BIC $\wh{\ell}(K)$, defined in \eqref{eq: dc-smbic}.
\end{itemize}
\item[2.] Calculate $\wh{K}= \argmax_{1\le K \le K_{\rm max}}\wh{\ell}(K)$.
\end{itemize}
\STATE \textbf{Output}: the optimal choice of the number of communities, $\wh{K}$.
\end{algorithmic}\label{alg: dcsbm}
\end{algorithm}

\begin{figure}[h!]
 \centering
 \includegraphics[width=1.0\linewidth]{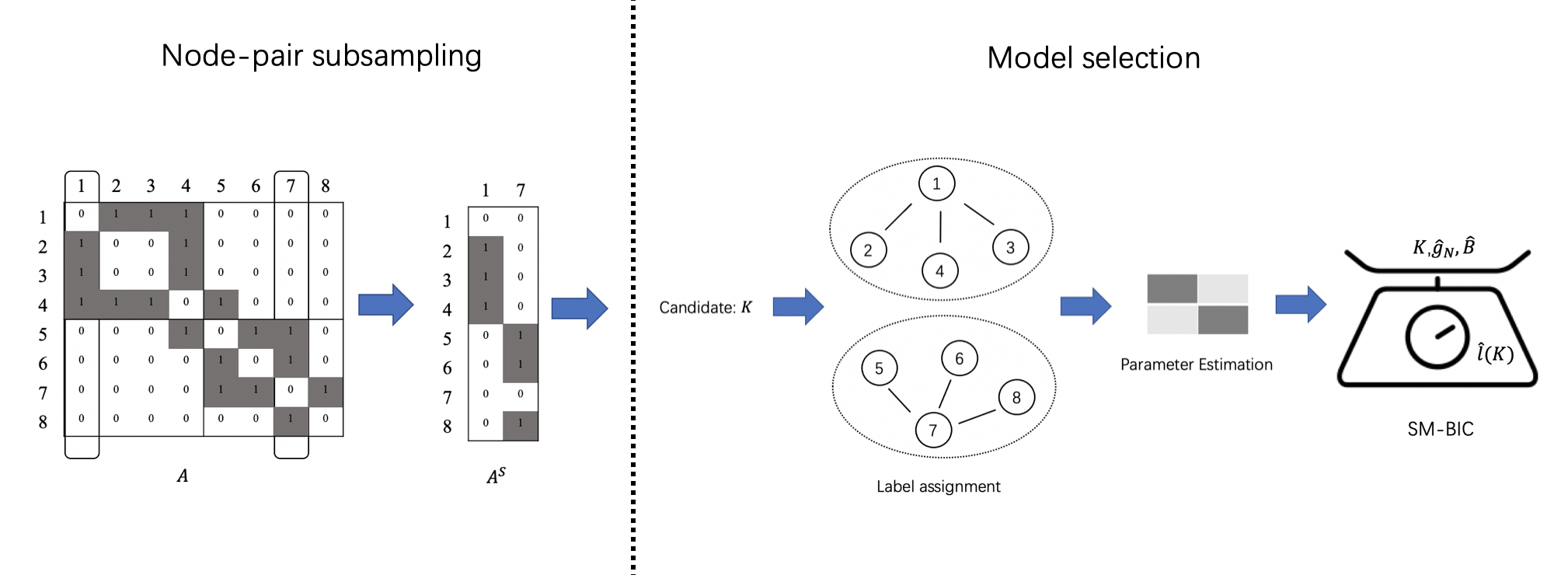}
 \caption{ An illustration of the steps to identify $K_0$ for SBM based on SM-BIC.}\label{fig: smbic}
\end{figure}

For convenience, we provide the model selection procedure for the SBM and DCSBM in Algorithms \ref{alg: sbm} and \ref{alg: dcsbm}, respectively. To illustrate the model selection algorithm, we show the procedure of identifying $K_0$ for SBM in Figure \ref{fig: smbic}. Moreover, based on the works of \citep{lei2015consistency, deng2021subsampling}, we demonstrated the consistency of spectral clustering for the sub-adjacency matrix $A^{\mS}$ in the supplementary materials. To show the effectiveness of the SM-BIC, we discuss its computational complexity in Proposition \ref{pro: computational}.
\begin{proposition}[\textbf{Computational complexity}]\label{pro: computational} Suppose that the subset of nodes $\mS$ is collected by simple random sampling $n$ nodes from $[N]$. Then, for both the SBM and DCSBM, the computational complexity of identifying $K_0$ by SM-BIC is $O(Nn)$.
\end{proposition}
 \noindent The proof of proposition \ref{pro: computational} is provided in Appendix \ref{appb: computational}. Note that for each candidate $K$, in the spectral clustering algorithm, we perform a truncated SVD to the sub-adjacency matrix, where the truncated SVD only computes the largest $K$ eigenvalues and the corresponding eigenvectors with computational complexity $O(Nn)$ for a constant $K$ \citep{feng2018faster, martin2018fast}. Proposition \ref{pro: computational} shows the computational advantage of the SM-BIC for large-scale networks. In the next section, we demonstrate that the required subsample size $n$ could be as small as $c(\log{N})^{2}$, where $c>0$ is a constant. In this case, the computational cost for identifying $K_0$ based on the SM-BIC could be $O\{N(\log{N})^2\}$.

\section{Theoretical properties}\label{sec: theoretical}

In this section, we discuss the theoretical properties of the SM-BIC. We first introduce some necessary conditions and subsequently discuss the required subsample size to ensure the effectiveness of the selected sample. Then, we demonstrate the consistency of the SM-BIC under the SBM and DCSBM. Namely, the criterion chooses the right $K_0$ with probability tending to one as $N$ goes to infinity.

\subsection{Basic assumptions and required subsampling size}

To discuss the theoretical properties of the SM-BIC, the following assumptions are considered.
\begin{enumerate}
\item (Network density) Assume $B^{*}= \rho_N \wt{B}^{*}$, where $\wt{B}^{*} \in (0,1)^{K_0 \times K_0}$ is a constant matrix and $\rho_N\to 0$ at a rate of $\rho_N N/\log{N} \to \infty.$ \label{ass: sparse}
\item (Balance level) There exists a constant $c>0$, such that $\min_{1\le k \le K_0} N_{k,g^{*}_N} \ge c N$. \label{ass: balance}
\end{enumerate}
\noindent Assumption \ref{ass: sparse} allows for sparse networks, where the network density is $\rho_N\to 0$ at the same rate as in the studies of \cite{wang2017likelihood}, \cite{hu2020corrected}, and \cite{li2020network}. Assumption \ref{ass: balance} requires the size of each community to be relatively balanced. This is a mild and common condition. For example, if the community assignment $g^{*}_N$ is generated from a multinomial distribution with parameters $\pi= (\pi_1, \cdots, \pi_{K_0})$ such that $\min_{1\le k\le K_0}\pi_k \ge c/{K_0}$, then Assumption \ref{ass: balance} is satisfied almost surely. This restriction is also used in \cite{lei2016goodness} and \cite{chen2018network}.

It is noteworthy that a small subsample leads to higher computational efficiency. However, if the subsample size is too small, it is difficult to guarantee the statistical validity of the proposed method. Therefore, we provide two necessary conditions to establish the lower bound of the subsample size $n$. First, we require that the subsampled nodes cover all blocks with high probability. Specifically, under ${\rm SBM}_{K_0}$, we define a set $\mathcal{M}_{K_0}= \{ \mS:   \forall \ k \in [K_0], \ \exists \ i \in \mS \ s.t., \ g^{*}_{N,i}=k\}$, where $g^{*}_{N,i}$ is the ground truth label of node $i$. This implies that the elements in $\mathcal{M}_{K_0}$ completely cover $K_0$ blocks. Second, we require that the average degree of the subnetwork should increase with $N$. Specifically, let $d_{i}= \sum_{j\in \mS}A_{ij}$ denote the degree of node $i$ in the subnetwork based on $A^{\mS}$, for $i=1,\cdots,N.$ Furthermore, let $d=\sum_{i=1}^{N}d_{i}/N$ denote the average degree of the subnetwork. Then, we assume the expected average degree $\mE(d)=\Omega(\log{N}).$ Based on these two conditions, we provide the lower bound of subsample size $n$ in the following proposition.

\begin{proposition}[\textbf{Subsample size}] \label{pro: size} Under Assumptions \ref{ass: sparse}--\ref{ass: balance}, suppose $\mS$ is collected by simple random sampling $n$ nodes from the entire network. If the subsample size is $n = \Omega(\log{N}/\rho_N)$, then we have $\mS \in \mathcal{M}_{K_0}$ and $\mE(d)=\Omega(\log{N})$ with high probability.
\end{proposition}
\noindent  The proof is provided in Appendix \ref{appb: size}. Note that $n = \Omega(\log{N}/\rho_N)$ means that there are positive constants $c$ and $N_0$ such that $n \ge c\log{N}/\rho_N$ for all $N> N_0$ \citep{knuth1976big}. According to Proposition \ref{pro: size}, the lower bound of subsample size goes to infinity with a lower speed compared to $N$. In particular, consider $\rho_N=(\log{N})^{-1}$, then the subsample size $n=\Omega\{(\log{N})^2\}$. Based on this proposition, we then demonstrate the consistency of this criterion.

\subsection{ Consistency of SM-BIC}

We first establish the consistency of the SM-BIC under SBMs. Given a subsampled adjacency matrix $A^{\mS}$, the underlying SM-BIC of ${\rm SBM}_{K_0}(g^{*}_N,B^{*})$ is
\beqrs
\ell^{*}(K_0)=  \log{f(A^{\mS}|g^{*}_N, B^{*})}- \left\{n\log{K_0} + \frac{ K_0(K_0+1)}{4}\log{M} \right\}.
\eeqrs
Intuitively, fitting the observed network with a correct number of communities yields the largest value of the SM-BIC. Then, for any candidate ${\rm SBM}_K$, we compare its SM-BIC $\ell(K)$ with the underlying SM-BIC $\ell^{*}(K_0)$ under three different cases, namely, under-fitting ($K<K_0$), correctly fitting ($K=K_0$), and over-fitting ($K>K_0$). We analyze the divergence between $\ell(K)$ and $\ell^{*}(K_0)$, which is
\beqr
\ell(K)- \ell^{*}(K_0)&=& \Big\{\max_{g_N\in \mC(A^{\mS},K)}\sup_{B\in \mB_{K}}\log{f(A^{\mS}|g_N,B)} - \log{f( A^{\mS}|g^{*}_N, B^{*})} \Big\}\nonumber\\
&-& \Big\{  n \log{(K/K_0)}+ \frac{K(K+1)- K_0(K_0+1)}{4}\log{M} \Big\}\nonumber\\
&=& L_{K,K_0}-R_{K,K_0},\label{eq: divided}
\eeqr
\noindent
where $L_{K,K_0}=\max_{g_N\in \mC(A^{\mS},K)}\sup_{B\in \mB_{K}}\log{f(A^{\mS}|g_N,B)} - \log{f( A^{\mS}|g^{*}_N, B^{*})}$, and $R_{K,K_0}= n \log{(K/K_0)}+ \{K(K+1)- K_0(K_0+1)\}/4\log{M}$. It is noteworthy that $L_{K,K_0}$ is a log-likelihood ratio, which measures the goodness-of-fit of the estimated model compared with the underlying model. Since $R_{K,K_0}$ is fixed for a given $K$ and $n$, we focus on analyzing $L_{K,K_0}$ in the three cases mentioned above.

\noindent
\textbf{ Case 1: Under-fitting.}  In this case, we prove the upper bound for the log-likelihood ratio $L_{K,K_0}$ in the following theorem.
\begin{theorem}[\textbf{Upper bound of the log-likelihood ratio under under-fitting}]\label{the: under} Suppose $A$ is generated from ${\rm SBM}_{K_0}(g^{*}_N,B^{*})$. Furthermore, suppose Assumptions \ref{ass: sparse}--\ref{ass: balance} hold and $n$ satisfies the condition in Proposition \ref{pro: size}. If $K<K_0$, then $L_{K,K_0} = -\Omega_P(\rho_NM).$
\end{theorem}
\noindent The technical proof of Theorem \ref{the: under} can be found in Appendix \ref{appc: under}. For $K<K_0$, it can be verified that $R_{K,K_0}= -\Omega(n+ \log{M})$. Combining the conclusion in Theorem \ref{the: under}, we have $\ell(K)- \ell^{*}(K_0)= - \Omega_P(\rho_NM)$ by \eqref{eq: divided}.  Moreover, note that the lower bound of the ratio $L_{K,K_0}$ is negatively related to $\rho_N$ and $M$, and goes to negative infinity as $N \to \infty$. This indicates that under the proposed conditions, the SM-BIC avoids the under-fitting case with high probability.

\noindent
\textbf{Case 2: Correctly fitting.} We then analyze the log-likelihood ratio $L_{K,K_0}$ under a given correct number of communities, i.e., $K=K_0$.
\begin{theorem}[\textbf{Convergence of the log-likelihood ratio under SBM}]\label{the: likelihood} Make the same assumptions as in Theorem \ref{the: under}. If $K=K_0$, then we have $L_{K_0,K_0}=O_P(\rho_N)$.
\end{theorem}
\noindent The proof is provided in Appendix \ref{appc: likelihood}. When $K=K_0$, since $R_{K_0,K_0}=0$, together with the conclusion in Theorem \ref{the: likelihood}, we have $\ell(K)- \ell^{*}(K_0)=L_{K_0,K_0}=O_P(\rho_N)$. Moreover, in Case 2, Theorem \ref{the: likelihood} implies that the log-likelihood ratio $L_{K_0,K_0}$ converges faster in sparse networks.

\noindent
\textbf{Case 3: Over-fitting.} Similar to the conclusion in Case 1, we present the upper bound of the log-likelihood ratio $L_{K,K_0}$ in the following theorem.
\begin{theorem}[\textbf{Upper bound of log-likelihood ratio under over-fitting}]\label{the: over} Make the same assumptions as in Theorem \ref{the: under}. For any candidate $K > K_0$, we have $L_{K,K_0}=O_P(\log{N}).$
\end{theorem}
\noindent The proof of this theorem can be found in Appendix \ref{appc: over}. For $K>K_0$, by the definition of $R_{K,K_0}$, we have  $R_{K,K_0}=\Omega(n+ \log{M})$. Then, together with the conclusion in Theorem \ref{the: over}, we have $\ell(K)-\ell^{*}(K_0)=- \Omega_P(n+ \log{M})$. Note that the upper bound is negatively related to the subsample size $n$, which indicates that the SM-BIC avoids over-fitting with increasing probability as $n$ grows. Therefore, Theorem \ref{the: over} ensures that the subsample size $n= \Omega(\log{N}/\rho_N)$ is large enough to prevent this misspecification.

To summarize, we establish the consistency of the SM-BIC under the SBM in the following corollary.
\begin{corollary}[\textbf{Consistent results for SBM}]\label{cor: consistency} Suppose $A$ is generated from ${\rm SBM}_{K_0}(g^{*}_N,B^{*})$ and Assumptions \ref{ass: sparse} and \ref{ass: balance} hold. If the subsample size $n$ satisfies the condition in Proposition \ref{pro: size}, then for $K\neq K_0$, we have $P(\ell(K)>\ell^{*}(K_0)) \to 0$, with $N \to \infty$.
\end{corollary}
\noindent Corollary \ref{cor: consistency} demonstrates that for the SBM, the correct number of communities can be identified by the SM-BIC with high probability.

Now, we investigate the consistency of the SM-BIC under the DCSBM. We assume that the degree heterogeneity parameter $\psi$ is known, which is also considered in the theoretical studies of \cite{lei2016goodness} and \cite{gao2018community}. In this case, according to criterion \eqref{eq: dc-smbic}, we have $L_{K,K_0}=\max_{g_N\in \mC(A^{\mS},K)}\sup_{B\in \mB_K} \log{f(A^{\mS}| g_N, B, \psi^{*})}-\log{f(A^{\mS}| g^{*}_N, B^{*}, \psi^{*})}.$ Then, we first investigate the convergence of the log-likelihood ratio under the correct specification.
\begin{theorem}[\textbf{Convergence of the log-likelihood ratio under DCSBM}]\label{the: convergence} Suppose that $A$ is generated from ${\rm DCSBM}_{K_0}(g^{*}_N,B^{*}, \psi^{*})$. Under Assumptions \ref{ass: sparse} and \ref{ass: balance}, if $n$ satisfies the condition in Proposition \ref{pro: size}, for $K=K_0$, we have $L_{K,K_0} =O_P(\rho_N)$.
\end{theorem}
\noindent The proof of Theorem \ref{the: convergence} is provided in Appendix \ref{appc: convergence}. According to Theorem \ref{the: convergence}, under the DCSBM, the convergence of $L_{K,K_0}$ can also be guaranteed if $K$ is correctly specified.

Based on Theorem \ref{the: convergence}, together with similar arguments, one can show that the conclusions of Theorem \ref{the: under} and Theorem \ref{the: over} hold under the DCSBM. Hence, we draw the theoretical results for the DCSBM as follows.
\begin{corollary}[\textbf{Consistent results for DCSBM}]\label{cor: dcsbm} Suppose $A$ is generated from ${\rm DCSBM}_{K_0}(g^{*}_N,B^{*},\psi^{*})$ and Assumptions \ref{ass: sparse}--\ref{ass: balance} hold. If $n$ satisfies the condition in Proposition \ref{pro: size}, for $K\neq K_0$, we have $P(\ell(K)>\ell^{*}(K_0)) \to 0$, with $N \to \infty$.
\end{corollary}

\section{Numerical studies}\label{sec: numerical}

\subsection{Simulation models and performance measurements}

We start with the generation mechanism of the networks. For a given $K_0$, we assume that the underlying node labels are generated by $g_{N,i}^{*} \sim {\rm Multinomial}(\pi)$ independently for all $i=1,\cdots, N$, where $\pi=(1/K_0, \cdots, 1/K_0)$. Second, we define the connectivity matrix as $B^{*}=\rho_N(\beta \one_{K_0}\one_{K_0}^{\top} + (1-\beta) I_{K_0})$, where $\one_{K_0} \in \mR^{K_0}$ is filled with elements 1 and $I_{K_0} \in \mR^{K_0\times K_0}$ is an identity matrix, and the {\it out-in-ratio} parameter $\beta\in (0,1)$ measures the connectivity divergence within and between communities.

Then, we evaluate the performance of the SM-BIC through the following three different examples under SBM framework.

\begin{example}[\textbf{Consistency of the approximated SM-BIC}]\label{exam: consistency} Let the number of communities $K_{0}$ vary from 2 to 5. For each $K_0$, let $N$ increase from 500 to 5,000. Furthermore, set the out-in-ratio parameter $\beta=0.15$ and let the network density $\rho_N=N^{-1/2}$. Then, according to Proposition \ref{pro: size}, take the subsample size as $n= \lceil \zeta\log{N}/\rho_N\rceil$, where $\lceil x \rceil$ represents the smallest integer of no less than $x$, and $\zeta$ is set to 1.0, 1.5, and 2.0, respectively.
\end{example}

\begin{example}[\textbf{The effect of network density}]\label{exam: density} Let the number of communities $K_0$ vary from 2 to 5 and the entire network size $N$ increase from 1,000 to 5,000. Additionally, take the out-in-ratio parameter as $\beta=0.15$ and let the network density $\rho_N$ increase from $0.5N^{-1/2}$ to $1.5N^{-1/2}$. For each network setting, we take the subsample size as $n= \lceil1.5N^{1/2}\log{N}\rceil$.
\end{example}

\begin{example}[\textbf{The effect of arbitrary outlier nodes}]\label{exam: outlier} According to the generalized stochastic block model proposed by \cite{cai2015robust}, we generate networks with a portion of outlier nodes. Specifically, assume that there are $N$ normal nodes and $m$ outlier nodes. The connections between $N$ normal nodes obey the ${\rm SBM}_{K_0}(g_{N}^{*},B^{*})$ with $\beta=0.15$ and $\rho_N=N^{-1/2}$, while connections between outliers are generated from a random graph model with a connectivity probability of 0.1. Moreover, define $X$ as a $N\times m$ matrix with independent Bernoulli entries, representing the connections between normal nodes and outlier nodes. Let $\mE X={\bm v} \one_{m}^{\top}$ where the components of ${\bm v}$ are $N$ i.i.d. copies of ${\bm u}^{2}/10$ and ${\bm u}$ is a uniform random variable on $[0,1]$. Furthermore, let $m$ increase from 20 to 100.
\end{example}

To further evaluate the performance of the SM-BIC method, we compare it with four existing approaches, namely, the method based on the Bethe Hessian matrix with moment correction (BHMC) proposed by \cite{le2015estimating}, the network cross-validation (NCV) method proposed by \cite{chen2018network}, the network cross-validation method by edge sampling (ECV) proposed by \cite{li2020network}, and the corrected Bayesian information criterion (CBIC) proposed by \cite{hu2020corrected}.

\begin{example}[\textbf{Comparison under SBM}]\label{exam: sbm} We generated the network from ${\rm SBM}_{K_0}(g_{N}^{*},B^{*})$ with $\beta=0.2$ and $\rho_N=N^{-1/2}$. Furthermore, let the network size $N$ increase from 3,000 to 5,000 and $K_0$ vary from 2 to 6, accordingly.
\end{example}

\begin{example}[\textbf{Comparison under DCSBM}]\label{exam: dcsbm} We follow the scenario proposed in \cite{zhao2012consistency}. The parameters $\psi_i$ are independently generated from a distribution with expectation 1, specifically,
\beqrs
\psi_i= \left\{
\begin{aligned}
\eta_i, \ \  &{\rm with \ probability } \ \alpha;\\
1/3, \ \ &{\rm with \ probability} \ (1-\alpha)/2;\\
5/3, \ \ &{\rm with \ probability} \ (1-\alpha)/2,
\end{aligned}
\right.
\eeqrs
where $\eta_i$ is uniformly distributed on the interval $[3/5, 7/5]$. The variance of $\psi_i$ is equal to $4\alpha/75+4(1-\alpha)/9$. Then, the variance is a decreasing function of $\alpha$. We vary $\alpha$ from 0.4 to 0.8. The other parameters are set to be the same in Example \ref{exam: sbm}.
\end{example}

Throughout this simulation study, we set the maximum candidate to $K_{\rm max}=10$. The random experiments are repeated $T=100$ times to ensure a reliable evaluation. Additionally, for each repetition, we assume the selected number of communities is $\wh{K}_t$, for $t=1,\cdots, T.$ Then, to gauge the performance of the SM-BIC, we consider two measurements.
First, the probability of correct identification is defined as ${\rm Prob}= \sum_{t=1}^{T} \bI(\wh{K}_t =K_0)/T,$ where a larger $\rm Prob$ corresponds to more accurate model selection.
 Second, the average of the selected number of communities is defined by ${\rm Mean} = \sum_{t=1}^{T} \wh{K}_t/T$. All simulations are conducted in a Linux server with a 3.60 GHz Intel Core i7-9700K CPU and 16 GB RAM.

\subsection{ Simulation results}

\begin{figure}[]
 \centering
 \includegraphics[width=\textwidth]{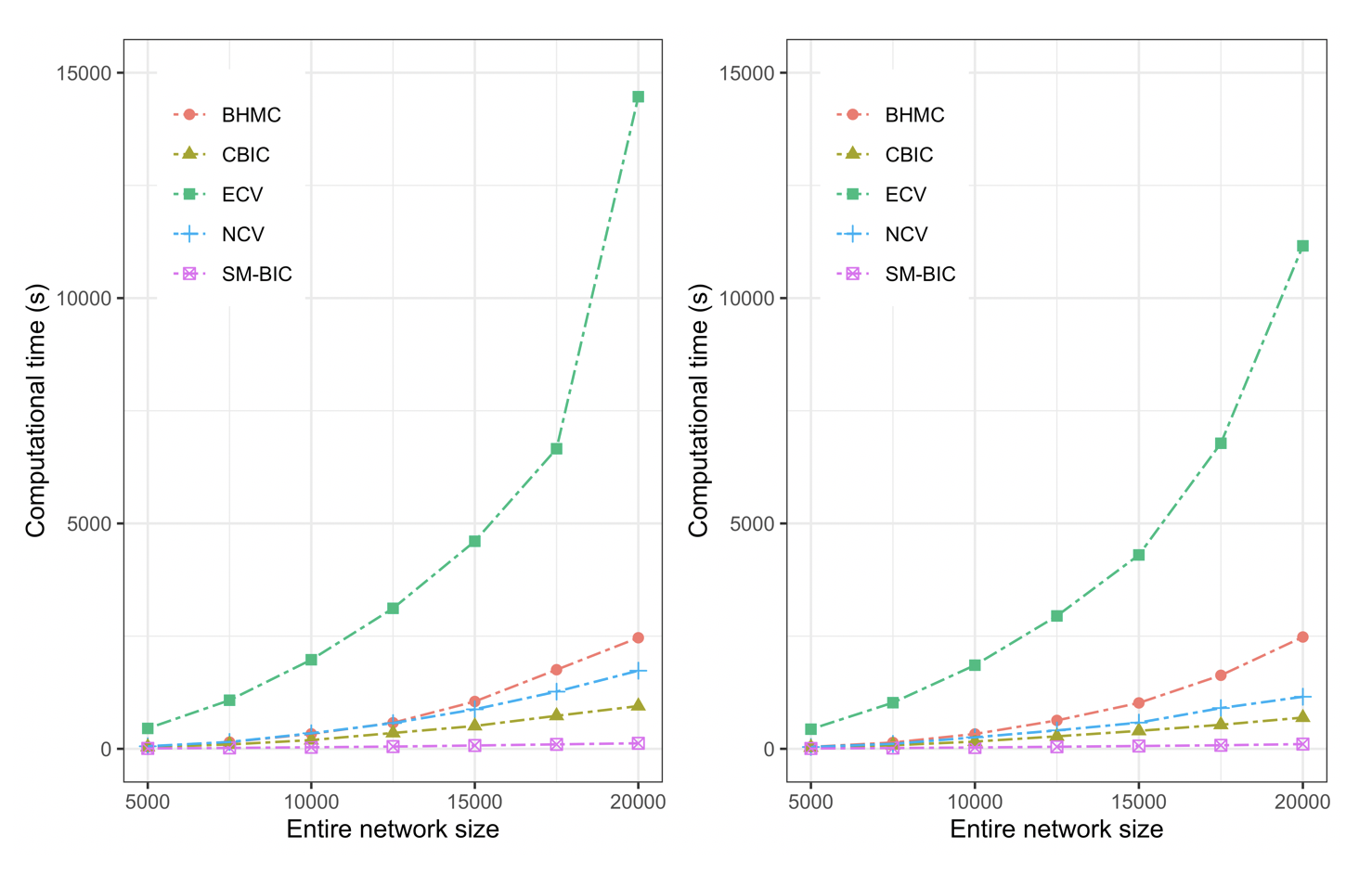}
\caption{Entire network size versus computational time (s) under the setting of Example \ref{exam: sbm}. The average CPU time (in seconds) is reported for two simulation settings: $K_0=2$ (left panel) and $K_0=6$ (right panel). Model selection methods are marked with different types of points.}\label{fig: cpu}
\end{figure}

All simulation results are shown in Tables \ref{tab: consistency}--\ref{tab: dcsbm} and Figure \ref{fig: cpu}. We draw the following conclusions from different examples.

{\sc Example \ref{exam: consistency}.} The simulation results are presented in Table \ref{tab: consistency}. We make the following comments. First, as $n$ grows from $\lceil \log{N}/\rho_N\rceil$ to $\lceil 2 \log{N}/\rho_N\rceil$, the probability of correct identification increases from 0.84 to 1.00 under the setting $K_0=5$ and $N=500$. Second, as $N$ increases from 500 to 5,000, the probability of correct identification increases from 0.84 to 1.00 under the setting $K_0=5$ and $n=\lceil \log{N}/\rho_N\rceil$. Third, as the network size $N$ increases from 500 to 5,000, the average CPU computational time of each experiment does not exceed 10.70 seconds. Hence, the SM-BIC is an efficient and consistent method for large-scale networks, and these results are consistent with our theoretical results in Proposition \ref{pro: size} and Corollary \ref{cor: consistency}.

\begin{table}[]
\centering
 \caption{ Simulation results of SM-BIC of Example \ref{exam: consistency}. The network density $\rho_N=N^{-1/2}$ and the network subsample size $n=\lceil \zeta\log{N}/\rho_N\rceil$. The measurements are provided and the average CPU computational time is also reported.}\label{tab: consistency}
\begin{tabular}{ll |ccr| ccr| ccr}
\hline
  &      & \multicolumn{3}{c|}{$N=500$} & \multicolumn{3}{c|}{$N=2,000$} & \multicolumn{3}{c}{$N=5,000$} \\
\hline
$K_0$ & $\zeta$ & Prob & Mean & CPU  & Prob    & Mean    & CPU    & Prob    & Mean   & CPU     \\
\hline
\multirow{3}{*}{2} & 1.0  & 1.00    & 2.00   & 0.73   & 1.00    & 2.00    & 2.08   & 1.00    & 2.00   & 6.77    \\
  & 1.5  & 1.00    & 2.00   & 0.76   & 1.00    & 2.00    & 2.41   & 1.00    & 2.00   & 8.51    \\
  & 2.0  & 1.00    & 2.00   & 0.78   & 1.00    & 2.00    & 2.71   & 1.00    & 2.00   & 10.70   \\
\hline
\multirow{3}{*}{3} & 1.0  & 1.00    & 3.00   & 0.69   & 1.00    & 3.00    & 1.93   & 1.00    & 3.00   & 6.36    \\
  & 1.5  & 1.00    & 3.00   & 0.71   & 1.00    & 3.00    & 2.19   & 1.00    & 3.00   & 7.97    \\
  & 2.0  & 1.00    & 3.00   & 0.72   & 1.00    & 3.00    & 2.45   & 1.00    & 3.00   & 9.93    \\
\hline
\multirow{3}{*}{4} & 1.0  & 0.96    & 4.00   & 0.65   & 1.00    & 4.00    & 1.82   & 1.00    & 4.00   & 6.14    \\
  & 1.5  & 0.99    & 4.01   & 0.67   & 1.00    & 4.00    & 2.03   & 1.00    & 4.00   & 7.67    \\
  & 2.0  & 1.00    & 4.00   & 0.67   & 1.00    & 4.00    & 2.26   & 1.00    & 4.00   & 9.55    \\
\hline
\multirow{3}{*}{5} & 1.0  & 0.84    & 4.86   & 0.62   & 1.00    & 5.00    & 1.73   & 1.00    & 5.00   & 5.99    \\
  & 1.5  & 1.00    & 5.00   & 0.64   & 1.00    & 5.00    & 1.95   & 1.00    & 5.00   & 7.49    \\
  & 2.0  & 1.00    & 5.00   & 0.65   & 1.00    & 5.00    & 2.16   & 1.00    & 5.00   & 9.20   \\
  \hline
\end{tabular}
\end{table}

\begin{table}[]
\centering
 \caption{ Simulation results of Example \ref{exam: density}. For each network setting, the network subsample size is set to $n= \lceil1.5N^{1/2}\log{N}\rceil$. The measurements are provided and the average CPU computational time is also reported.}\label{tab: density}
\begin{tabular}{ll |ccr |ccr| ccr}
\hline
  &      & \multicolumn{3}{c|}{$N=1,000$} & \multicolumn{3}{c|}{$N=3,000$} & \multicolumn{3}{c}{$N=5,000$} \\
\hline
$\rho_N N^{1/2}$                  & $K_0$    & Prob    & Mean    & CPU     & Prob    & Mean    & CPU     & Prob    & Mean    & CPU     \\
\hline
\multirow{4}{*}{0.5} & 2 & 1.00    & 2.00    & 1.24    & 1.00    & 2.00    & 3.74    & 1.00    & 2.00    & 8.41    \\
                     & 3 & 1.00    & 3.00    & 1.16    & 1.00    & 3.00    & 3.56    & 1.00    & 3.00    & 8.02    \\
                     & 4 & 0.99    & 3.99    & 1.13    & 1.00    & 4.00    & 3.41    & 1.00    & 4.00    & 7.99    \\
                     & 5 & 0.75    & 4.75    & 1.11    & 1.00    & 5.00    & 3.28    & 1.00    & 5.00    & 7.73    \\
\hline
\multirow{4}{*}{1.0} & 2 & 1.00    & 2.00    & 1.26    & 1.00    & 2.00    & 3.88    & 1.00    & 2.00    & 8.63    \\
                     & 3 & 1.00    & 3.00    & 1.15    & 1.00    & 3.00    & 3.53    & 1.00    & 3.00    & 8.04    \\
                     & 4 & 1.00    & 4.00    & 1.08    & 1.00    & 4.00    & 3.31    & 1.00    & 4.00    & 7.71    \\
                     & 5 & 1.00    & 5.00    & 1.03    & 1.00    & 5.00    & 3.19    & 1.00    & 5.00    & 7.59    \\
\hline
\multirow{4}{*}{1.5} & 2 & 1.00    & 2.00    & 1.21    & 1.00    & 2.00    & 3.92    & 1.00    & 2.00    & 8.85    \\
                     & 3 & 1.00    & 3.00    & 1.08    & 1.00    & 3.00    & 3.48    & 1.00    & 3.00    & 7.96    \\
                     & 4 & 1.00    & 4.00    & 1.00    & 1.00    & 4.00    & 3.25    & 1.00    & 4.00    & 7.56    \\
                     & 5 & 1.00    & 5.00    & 0.95    & 1.00    & 5.00    & 3.07    & 1.00    & 5.00    & 7.33   \\
\hline
\end{tabular}
\end{table}

{\sc Example \ref{exam: density}.} The simulation results are provided in Table \ref{tab: density}. We obtain the following findings. First, as network density $\rho_N$ increases from $0.5N^{-1/2}$ to $1.5N^{-1/2}$, the probability of correct identification increases to 1 for all $K_0=2,\cdots,5.$ Second, even in the sparsest case $\rho_N= 0.5 N^{-1/2}$, as $N$ grows from 1,000 to 5,000, the probability of correct identification increases from 0.75 to 1.00. Hence, for large-scale networks, the proposed method allows for a higher level of sparsity.

{\sc Example \ref{exam: outlier}.} This simulation results are provided in Table \ref{tab: outlier}. We draw the following conclusions. First, as the number of outliers decreases from 100 to 20, the accuracy of recovering $K_0$ increases from 0.82 to 1.00 under the setting $N=2,000$ and $K_0=5$. Second, as $N$ varies from 2,000 to 5,000, the probability of correct identification grows from 0.82 to 1.00 in the case of $K_0=5$. Therefore, for large-scale networks with arbitrary outliers, the SM-BIC method can accurately identify the number of communities with high probability.

\begin{table}[]
\centering
 \caption{Simulation results of Example \ref{exam: outlier}. In this study, the network density is $\rho_N=N^{-1/2}$ and the subsample size is $n= \lceil 1.5\log{N}/\rho_N \rceil$. Furthermore, for each network with $N$ nodes, the number of outlier nodes $m$ increases from 20 to 100. The measurements are provided and the average CPU computational time is also reported.}\label{tab: outlier}
\begin{tabular}{ll |ccr |ccr |ccr}
\hline
                     &   & \multicolumn{3}{c|}{$N=2,000$} & \multicolumn{3}{c|}{$N=3,000$} & \multicolumn{3}{c}{$N=5,000$} \\
 \hline
$m$                    & $K_0$ & Prob    & Mean    & CPU     & Prob    & Mean    & CPU     & Prob    & Mean    & CPU     \\
\hline
\multirow{4}{*}{20}  & 2 & 1.00    & 2.00    & 2.36    & 1.00    & 2.00    & 3.79    & 1.00    & 2.00    & 7.80    \\
                     & 3 & 1.00    & 3.00    & 2.15    & 1.00    & 3.00    & 3.52    & 1.00    & 3.00    & 7.34    \\
                     & 4 & 1.00    & 4.00    & 2.00    & 1.00    & 4.00    & 3.15    & 1.00    & 4.00    & 7.26    \\
                     & 5 & 1.00    & 5.00    & 1.89    & 1.00    & 5.00    & 3.00    & 1.00    & 5.00    & 7.02    \\
\hline
\multirow{4}{*}{50}  & 2 & 1.00    & 2.00    & 2.39    & 1.00    & 2.00    & 3.71    & 1.00    & 2.00    & 7.59    \\
                     & 3 & 1.00    & 3.00    & 2.17    & 1.00    & 3.00    & 3.44    & 1.00    & 3.00    & 7.41    \\
                     & 4 & 1.00    & 4.00    & 2.00    & 1.00    & 4.00    & 3.06    & 1.00    & 4.00    & 7.24    \\
                     & 5 & 1.00    & 5.00    & 1.87    & 1.00    & 5.00    & 2.89    & 1.00    & 5.00    & 6.88    \\
\hline
\multirow{4}{*}{100} & 2 & 1.00    & 2.00    & 2.42    & 1.00    & 2.00    & 3.59    & 1.00    & 2.00    & 7.79    \\
                     & 3 & 1.00    & 3.00    & 2.19    & 1.00    & 3.00    & 3.41    & 1.00    & 3.00    & 7.62    \\
                     & 4 & 0.97    & 4.03    & 2.02    & 1.00    & 4.00    & 3.20    & 1.00    & 4.00    & 7.30    \\
                     & 5 & 0.82    & 5.18    & 1.89    & 0.99    & 5.01    & 2.95    & 1.00    & 5.00    & 6.90   \\
\hline
\end{tabular}
\end{table}

{\sc Example \ref{exam: sbm}.} The comparison results are shown in Table \ref{tab: sbm} and Figure \ref{fig: cpu}. We draw the following conclusions. First, SM-BIC is more accurate than the ECV method in this study. Specifically, for the setting of $N=5,000$, when $K_0=4$ and $K_0=6$, the Prob of the ECV method is only 0.83 and 0.84, respectively, while the Prob of the SM-BIC is 1.00 in these cases.
Second, the average computational time of the SM-BIC is much smaller than that of the BHMC, NCV, and CBIC, especially when $N$ is large. As shown in Figure \ref{fig: cpu}, the average CPU computational time of these methods is further compared across diverse network sizes. We observe that the average CPU computational time of the SM-BIC is the smallest, while the ECV method is much more computationally expensive than other algorithms. Because in each iteration, ECV performs matrix completion and estimates community labels from a $N\times N$-dimensional low-rank matrix.

\begin{table}[]
\centering
 \caption{Simulation results of Example \ref{exam: sbm}. The network density $\rho_N=N^{-1/2}$ and the subsample size $n= \lceil 1.5\log{N}/\rho_N\rceil$.The measurements of each method are provided and the average CPU computational time is also reported.}\label{tab: sbm}
\begin{tabular}{ll| ccr| ccr| ccr}
\hline
                       &        & \multicolumn{3}{c|}{$K_0=2$} & \multicolumn{3}{c|}{$K_0=4$} & \multicolumn{3}{c}{$K_0=6$} \\
\hline
$N$           &    Method    & Prob  & Mean  & CPU     & Prob  & Mean  & CPU     & Prob  & Mean  & CPU     \\
\hline
\multirow{5}{*}{3,000} & BHMC   & 1.00  & 2.00  & 8.99    & 1.00  & 4.00  & 9.73    & 1.00  & 6.00  & 9.18    \\
                       & NCV    & 1.00  & 2.00  & 15.23   & 0.99  & 4.01  & 13.39   & 0.94  & 6.08  & 10.89   \\
                       & ECV    & 1.00  & 2.00  & 156.13  & 0.92  & 4.08  & 153.35  & 0.84  & 6.22  & 148.75  \\
                       & CBIC   & 1.00  & 2.00  & 13.18   & 1.00  & 4.00  & 12.04   & 1.00  & 6.00  & 10.46   \\
                       & SM-BIC & 1.00  & 2.00  & 3.88    & 1.00  & 4.00  & 3.57    & 1.00  & 6.00  & 3.16    \\
\hline
\multirow{5}{*}{4,000} & BHMC   & 1.00  & 2.00  & 22.14   & 1.00  & 4.00  & 22.17   & 1.00  & 6.00  & 22.69   \\
                       & NCV    & 1.00  & 2.00  & 34.91   & 1.00  & 4.00  & 27.70   & 0.94  & 6.06  & 25.10   \\
                       & ECV    & 1.00  & 2.00  & 284.49  & 0.91  & 4.12  & 272.21  & 0.83  & 6.22  & 272.80  \\
                       & CBIC   & 1.00  & 2.00  & 24.94   & 1.00  & 4.00  & 21.34   & 1.00  & 6.00  & 19.80   \\
                       & SM-BIC & 1.00  & 2.00  & 5.99    & 1.00  & 4.00  & 5.44    & 1.00  & 6.00  & 5.08    \\
\hline
\multirow{5}{*}{5,000} & BHMC   & 1.00  & 2.00  & 42.88   & 1.00  & 4.00  & 43.13   & 1.00  & 6.00  & 42.91   \\
                       & NCV    & 1.00  & 2.00  & 58.04   & 0.98  & 4.04  & 48.98   & 0.93  & 6.09  & 40.00   \\
                       & ECV    & 1.00  & 2.00  & 460.89  & 0.83  & 4.25  & 444.04  & 0.84  & 6.22  & 437.66  \\
                       & CBIC   & 0.98  & 2.04  & 40.44   & 1.00  & 4.00  & 36.21   & 1.00  & 6.00  & 31.52   \\
                       & SM-BIC & 1.00  & 2.00  & 9.20    & 1.00  & 4.00  & 8.47    & 1.00  & 6.00  & 7.55   \\
\hline
\end{tabular}
\end{table}

{\sc Example \ref{exam: dcsbm}.}  The comparison results are reported in Table \ref{tab: dcsbm}. We draw the following conclusions. First, as $\alpha$ decreases from 0.8 to 0.4, the accuracy of the ECV method decreases from 1.00 to 0.00 under the settings $K_0=4, 5, 6$. However, the SM-BIC method can correctly identify $K_0$ in these cases. Second, compared with the BHMC, NCV, ECV, and CBIC methods, when $\alpha=0.4$ and $K_0=2$, the average CPU computational time of the BHMC, NCV, ECV, and CBIC methods is 41.88s, 86.19s, 482.43s, and 68.75s, respectively, while that of the SM-BIC method is only 11.83s.
In this way, in the DCSBM, the SM-BIC is more robust than the ECV method in terms of degree heterogeneity and more computationally efficient than all these methods.

\begin{table}[]
 \centering
 \caption{Simulation results of Example \ref{exam: dcsbm}. The network density $\rho_N=N^{-1/2}$ and the subsample size is $n= \lceil 1.5\log{N}/\rho_N\rceil$. Moreover, the heterogeneity parameter $\alpha$ varies from 0.4 to 0.8. The measurements of each method are provided and the average CPU computational time is also reported.}\label{tab: dcsbm}
\begin{tabular}{ll |rrr| rrr| rrr}
  \hline
                   &         & \multicolumn{3}{c|}{$\alpha=0.4$} & \multicolumn{3}{c|}{$\alpha=0.6$} & \multicolumn{3}{c}{$\alpha=0.8$} \\
  \hline
$K_0$            & Method & Prob    & Mean    & CPU       & Prob    & Mean    & CPU       & Prob    & Mean    & CPU       \\
  \hline
\multirow{5}{*}{2} & BHMC    & 1.00    & 2.00    & 41.88     & 1.00    & 2.00    & 43.42     & 1.00    & 2.00    & 42.43     \\
                   & NCV     & 1.00    & 2.00    & 86.19     & 1.00    & 2.00    & 84.60     & 1.00    & 2.00    & 87.78     \\
                   & ECV     & 1.00    & 2.00    & 482.43    & 1.00    & 2.00    & 485.45    & 1.00    & 2.00    & 486.20    \\
                   & CBIC    & 1.00    & 2.00    & 68.75     & 1.00    & 2.00    & 69.29     & 1.00    & 2.00    & 70.47     \\
                   & SM-BIC  & 1.00    & 2.00    & 11.83     & 1.00    & 2.00    & 11.75     & 1.00    & 2.00    & 11.69     \\
  \hline
\multirow{5}{*}{3} & BHMC    & 1.00    & 3.00    & 41.30     & 1.00    & 3.00    & 42.05     & 1.00    & 3.00    & 42.50     \\
                   & NCV     & 1.00    & 3.00    & 72.58     & 1.00    & 3.00    & 79.44     & 1.00    & 3.00    & 79.94     \\
                   & ECV     & 1.00    & 3.00    & 469.93    & 1.00    & 3.00    & 475.81    & 1.00    & 3.00    & 476.57    \\
                   & CBIC    & 1.00    & 3.00    & 62.52     & 1.00    & 3.00    & 64.16     & 1.00    & 3.00    & 64.55     \\
                   & SM-BIC  & 1.00    & 3.00    & 11.26     & 1.00    & 3.00    & 11.89     & 1.00    & 3.00    & 11.72     \\
  \hline
\multirow{5}{*}{4} & BHMC    & 1.00    & 4.00    & 41.48     & 1.00    & 4.00    & 42.37     & 1.00    & 4.00    & 42.23     \\
                   & NCV     & 1.00    & 4.00    & 70.65     & 1.00    & 4.00    & 70.64     & 1.00    & 4.00    & 71.72     \\
                   & ECV     & 0.00    & 5.10    & 468.62    & 1.00    & 4.00    & 467.08    & 1.00    & 4.00    & 470.19    \\
                   & CBIC    & 1.00    & 4.00    & 59.80     & 1.00    & 4.00    & 60.49     & 1.00    & 4.00    & 61.28     \\
                   & SM-BIC  & 1.00    & 4.00    & 11.03     & 1.00    & 4.00    & 11.20     & 1.00    & 4.00    & 11.09     \\
  \hline
\multirow{5}{*}{5} & BHMC    & 1.00    & 5.00    & 41.44     & 1.00    & 5.00    & 42.13     & 1.00    & 5.00    & 43.36     \\
                   & NCV     & 1.00    & 5.00    & 67.62     & 1.00    & 5.00    & 62.72     & 1.00    & 5.00    & 63.20     \\
                   & ECV     & 0.00    & 6.00    & 469.28    & 1.00    & 5.00    & 465.62    & 1.00    & 5.00    & 468.01    \\
                   & CBIC    & 1.00    & 5.00    & 57.91     & 1.00    & 5.00    & 56.64     & 1.00    & 5.00    & 57.57     \\
                   & SM-BIC  & 1.00    & 5.00    & 10.74     & 1.00    & 5.00    & 10.42     & 1.00    & 5.00    & 10.52     \\
  \hline
\multirow{5}{*}{6} & BHMC    & 1.00    & 6.00    & 42.08     & 1.00    & 6.00    & 43.38     & 1.00    & 6.00    & 42.89     \\
                   & NCV     & 1.00    & 6.00    & 61.47     & 1.00    & 6.00    & 58.75     & 1.00    & 6.00    & 61.01     \\
                   & ECV     & 0.00    & 7.20    & 465.68    & 0.00    & 7.25    & 464.23    & 1.00    & 6.00    & 462.87    \\
                   & CBIC    & 1.00    & 6.00    & 54.42     & 1.00    & 6.00    & 53.97     & 1.00    & 6.00    & 54.95     \\
                   & SM-BIC  & 1.00    & 6.00    & 10.40     & 1.00    & 6.00    & 10.21     & 1.00    & 6.00    & 10.25    \\
  \hline
\end{tabular}
\end{table}

\subsection{Real data analysis}

\noindent
\textbf{Political blog dataset.} The political blog dataset was collected and analyzed in \cite{adamic2005political}. The data set consists of over one thousand blogs discussing US politics, with edges representing web links. The nodes are labeled as being either ``conservative" or ``liberal", which can be treated as two well-defined communities. We only consider the largest connected component of this network, which consists of 1,222 nodes with community sizes of 586 and 636, while the network density is $\rho_N=2.24\%$. The degree-corrected stochastic block model is believed to fit better for this network than stochastic block model \citep{karrer2011stochastic, zhao2011community}. Then, under the DCSBM framework, we take the subsample size of the SM-BIC as $n=\lceil1.5\log{N}/\rho_N \rceil=475$, and compare the SM-BIC method with other algorithms. Specifically, we obtain the estimated number of communities as 2 by the NCV, CBIC, and SM-BIC, with computation times of 5.57s, 3.82s, and 1.60s, respectively. While the BHMC and ECV estimate $\wh{K}=7$ and $\wh{K}=6$, respectively. We see that the NCV, CBIC, and SM-BIC methods all give correct estimates for the number of communities, and SM-BIC further outperforms these two algorithms in terms of computational efficiency.

\noindent
\textbf{ A house price dataset.} This dataset is publicly available on the platform {\it Kaggle} (\url{https://Kaggle.com}), which contains housing transaction information in Beijing from 2011 to 2017. Here, we collect 6,000 samples traded in 2016, distributed in the ``Feng Tai", ``Chang Ping", and ``Hai Dian" districts of Beijing. The nodes are these collected samples and a network is obtained by randomly connecting the node pairs in the same district with a probability of 0.1. That is, if node $i$ and $j$ are in the same district, then we add an edge to node pair $(i, j)$ with probability 0.1.
As a result, this network has three well-defined communities with the sizes of communities 1,661, 2,365, and 1,974, respectively, while the network density is $\rho_N=3.40\%$.
We then apply the SM-BIC and the aforementioned methods to identify the number of communities for this network under the SBM and DCSBM frameworks, respectively. For the SM-BIC method, the subsample size is set to be $n=\lceil2\log{N}/\rho_N \rceil=511$. The results are provided in Table \ref{tab: house}. As shown in Table \ref{tab: house}, we observe that the SM-BIC method can correctly identify the number of communities under both the SBM and DCSBM frameworks. Moreover, the SM-BIC takes only 8.75s for SBM, which is only 11.0\% of BHMC, 12.4\% of NCV, and 1.4\% of ECV, respectively. For the DCSBM model, the SM-BIC takes 12.58s, which is only 16.3\% of BHMC, 11.3\% of NCV, 1.8\% of ECV, and 9.5\% of CBIC, respectively.

\begin{table}[]
\centering
 \caption{Comparison results of different methods in the dataset of housing prices in Beijing. The estimated number of communities $\wh{K}$ and the CPU computational time of each method are reported.}\label{tab: house}
\begin{tabular}{cl rrrrr}
\hline
\multicolumn{1}{l}{Model} &     & BHMC    & NCV    & ECV    & CBIC   & SMBIC \\
\hline
\multirow{2}{*}{SBM}       & $\wh{K}$   & 3.00  & 3.00   & 3.00   & 4.00   & 3.00  \\
                           & CPU & 79.21 & 70.50  & 646.21 & 52.77  & 8.75  \\
\\
\multirow{2}{*}{DCSBM}     & $\wh{K}$   & 3.00  & 3.00   & 3.00   & 3.00   & 3.00  \\
                           & CPU & 77.26 & 111.76 & 686.25 & 132.45 & 12.58\\
 \hline
\end{tabular}
\end{table}

\section{Concluding remarks}\label{sec: concluding}

This work proposes a subsampling-based modified Bayesian information criterion (SM-BIC) to identify the number of communities for large-scale SBMs. We also extend this criterion to DCSBMs. Specifically, the technical conditions of subsampling size are derived, and the consistency properties of SM-BIC are established. In the context of large-scale networks, the proposed SM-BIC has more valuable computational advantages than existing model selection methods. Specifically, the computational complexity of the SM-BIC for both the SBM and DCSBM could be as low as $O\{N(\log{N})^{2}\}$. Consequently, the SM-BIC method could be performed even using a personal computer. Numerical studies further demonstrate these computational improvements.

To conclude this work, we consider several interesting topics for future research. First, in this study, we focus on reducing computational costs by network subsampling only once; this idea can be extended to a resampling approach, which is currently under investigation. Second, informative subsamples are important for extracting useful information from the entire network. Subsampling strategies for independent big data have been extensively studied; see \cite{matias2019speeding}, \cite{wang2021optimal}, and \cite{yu2022optimal} for further discussions. Based on these studies, it would be interesting to investigate subnetwork extraction methods with meaningful statistical interpretations in large-scale networks. Third, in this work, following \cite{ma2021determining}, we assume that $K_0$ is fixed. However, it is an interesting and challenging question to allow for a diverging $K_0$. We will work in this direction in future research.

The code is publicly available on GitHub (\url{https://github.com/Stamath/SMBIC}).

\appendix
\section{Necessary notations and lemmas}\label{appa}

In Appendix \ref{appa}, we introduce some necessary notations in Appendix \ref{appa: notation}. Then, we give three useful lemmas for the subsequent theoretical proof of the proposed method in Appendix \ref{appa: lemma}.

\subsection{Notations}\label{appa: notation}

Given a label vector $g_{N}$, we define some necessary count statistics. Define a $K\times K$ count matrix as $n_{g_{N}}= (n_{kl,g_{N}})_{1\le k,l\le K}$ and $o_{g_{N}}= (o_{kl,g_{N}})_{1\le k,l\le K}$. Let ${\bm p}= (N_{1,g_{N}^{*}}, \cdots, N_{K,g_{N}^{*}})^{\top}/N$ denote the underlying block proportions, where $N_{k,g_{N}^{*}}= \sum_{i=1}^{N}\bI(g_{N,i}^{*}=k)$ represents the number of nodes belonging to the $k$-th cluster. For two sets of labels $g_{N}$ and $g'_{N}$, define $|g_{N}-g'_{N}|= \sum_{i=1}^{N}\bI(g_{N,i} \neq g'_{N,i})|$. In addition, define $\tau$ as a permutation on $[K]$ and denote $\|\cdot \|_{\infty}$ as a maximum norm of a matrix.

{\sloppy
For simplicity, we quote the notations from \cite{wang2017likelihood} to characterize the log-likelihood function. Let $H_{g_{N}}$ be an $K \times K_0$ confusion matrix whose $(k,l)$-entry is $H_{kl,g_{N}}=1/N\sum_{i=1}^{N} \bI{\{g_{N,i}=k, g^{*}_{N,i}=l\}}.$ Additionally, we define
 \beqrs
 F(Q, q)= \sum_{1\le k\le l \le K}q_{kl}\gamma\left(\frac{Q_{kl}}{q_{kl}}\right),
 \eeqrs
  where $\gamma(x)= x\log{x} + (1-x)\log(1-x)$ for $x\in (0,1)$. Then, for a fixed label vector $g_{N}$, the corresponding log-likelihood can be expressed as
 $\sup_{B\in \mB_K}\log{f(A^{\mS} |g_{N}, B)}= MF(o_{g_{N}}/M, n_{g_{N}}/M).$
 We further define its expectation as
 \beqrs
 G(H_{g_{N}}, B^{*})= \sum_{1\le k\le l\le K} ( H_{g_{N}} \one\one^{\top}H_{g_{N}}^{\top})_{kl} \gamma\Big\{\frac{ ( H_{g_{N}}B^{*}H_{g_{N}}^{\top})_{kl} }{ (H_{g_{N}} \one\one^{\top}H_{g_{N}}^{\top})_{kl}}\Big\}.
 \eeqrs
}

\subsection{Useful lemmas}\label{appa: lemma}

Here, we provide some useful lemmas, that is, Lemmas \ref{lem: hoeffding}--\ref{lem: concentration}, for the proof of the consistency of the SM-BIC.

In statistics, Hoeffding inequality provides an upper bound for the sum of bounded random variables, which was proved by \cite{hoeffding1963probability}.
\begin{lemma}[\textbf{Hoeffding inequality}]\label{lem: hoeffding} Let $x_i, i=1,\cdots, N,$ be mutually independent random variables such that $a_i \le x_i \le b_i$ almost surely. Consider the sum of these random variables, $Y_N= \sum_{i=1}^{N}x_i$. Then, for all $s >0$,
\beqrs
P\{ Y_N- E(Y_N) \ge s \} \le \exp{\Big\{-\frac{2s^{2}}{\sum_{i=1}^{N}(b_i -a_i)^{2}}\Big\}}.
\eeqrs
\end{lemma}

In the under-fitting case, without loss of generality, we start with $K=K_0-1$, and the following Lemma \ref{lem: expectation} shows that $G(H_{g_{N}}, B^{*})$ is maximized by combining two existing communities in $g_{N}^{*}$.
\begin{lemma}[\textbf{Expectation of the log-likelihood function of under-fitting}]\label{lem: expectation} Given the true label $g_{N}^{*}$, suppose $g_{N} \in \mC(A^{\mS},K_0-1)$, and then maximizing the function $G(H_{g_{N}}, B^{*})$ over $H_{g_N}$ achieves its maximum in the label set $$\{g_{N} \in \mC(A^{\mS},K_0-1): \ \text{there exists}\ \tau \ \text{such that} \ \tau(g_{N})= U_{kl}(g_{N}^{*}), 1\le k,l\le K_0\},$$
where $U_{k,l}(g_{N}^{*})$ merges $g_{N}^{*}$ with labels $k$ and $l$. Furthermore, suppose $g'_N$ gives the unique maximum (up to a permutation $\tau$), and for all $H_{g_{N}}$, there exists a positive constant $c_1>0$ such that $H_{g_{N}} \ge0$, $H^{\top}_{g_{N}}\one= {\bm p}$,
\beqrs
\frac{\partial G\{(1-\epsilon)H_{g'_N} +\epsilon H_{g_{N}}, B^{*}\}}{\partial \epsilon} \Bigg\vert_{\epsilon=0^+} < -c_1<0.
\eeqrs
\end{lemma}

For subsampled adjacency matrix $A^{\mS}$, consider $\|A^{\mS}\|_{\infty}= \max\limits_{1\le i\le N}\sum_{j\in \mS}|A_{ij}|$. The following Lemma \ref{lem: concentration} provides a concentration inequality to bound the variation in the adjacency matrix $A^{\mS}$, as proposed by \cite{wang2017likelihood}.
\begin{lemma}[\textbf{Concentration inequality}]\label{lem: concentration} Assume $g_{N} \in \mC(A^{\mS},K)$ and define $W_{g_{N}}= o_{g_{N}}/M- H_{g_{N}}B^{*}H_{g_{N}}^{\top}$. For $\epsilon \le 3$,
\beqrs
P\left\{ \max_{g_{N}\in \mC(A^{\mS},K)} \|W_{g_{N}}\|_{\infty} > \epsilon\right\} \le 2K^{N+2} \exp\{-c_1(B^{*})\epsilon^2\rho_N^{-1}M\},
\eeqrs
where $c_{1}(B^{*})$ is a constant depending on $B^{*}$ and $M=Nn-n(n+1)/2$. Let $\omega_n=(\rho_NN\log{n}/M)^{1/2}$, then $\max_{g_{N}\in \mC(A^{\mS},K)} \|W_{g_{N}}\|_{\infty} > \omega_n \to 0,$ with high probability, for $n, N\to \infty$. Furthermore, let $g'_N \in \mC(A^{\mS},K)$ be a fixed set of labels; then, for $\epsilon\le \frac{3m}{N}$,
\beqrs
\begin{split}
&P\Big(\max_{g_{N}: |g_{N}-g'_N|\le m} \| W_{g_{N}} -W_{g'_{N}}\|_{\infty} > \epsilon \Big)\\
& \le 2 \binom Nm K^{m+2} \exp{\Big\{ - c_2(B^{*}) \frac{N^{3}\epsilon^{2}}{\rho_Nm} \Big\} },
\end{split}
\eeqrs
where $m$ is an integer and $c_{2}(B^{*})$ is a constant depending on $B^{*}.$
\end{lemma}

 \section{Demonstrations of SM-BIC}\label{appb}

In Appendix \ref{appb}, we use the BIC approximation to prove Lemma \ref{lem: log-likelihood}, shown in Appendix \ref{appb: likelihood}. Furthermore, we provide the proofs of Propositions \ref{pro: computational} and \ref{pro: size} in Appendices \ref{appb: computational} and \ref{appb: size}, respectively.

\subsection{Proof of Lemma \ref{lem: log-likelihood}}\label{appb: likelihood}

The proof of the log-likelihood function approximation can be accomplished by the following two steps. First, we use Taylor approximation for the likelihood function, i.e., $f(A^{\mS}|g_{N})$. Then, we investigate its Hessian matrix.

{\sc Step 1.} Assume that the likelihood function $f(A^{\mS}|g_{N},\theta)$ attains its maximum at $\wh{\theta}$ so that $\partial f(A^{\mS}|g_{N},\theta)/\partial \theta|_{\theta= \wh{\theta}}=0$. By Taylor expansion, we have,
\beqrs
\log{f(A^{\mS}|g_{N},\theta)} \approx \log{f(A^{\mS}| g_{N},\wh{\theta})}+ \frac{1}{2}(\theta- \wh{\theta})^{\top}D(\theta- \wh{\theta}),
\eeqrs
where $D$ is a $\frac{K(K+1)}{2}\times \frac{K(K+1)}{2}$ matrix such that for $1\le k' , l' \le \frac{K(K+1)}{2}$, $$D_{k'l'}= \frac{\partial^{2}f(A^{\mS}|g_{N},\theta)}{\partial\theta_{k'}\partial \theta_{l'}}.$$ Since $f(A^{\mS}|g_{N},\theta)$ attains its maximum at $\wh{\theta}$, the Hessian matrix $D$ is negative definite. Let $\wt{D}= -D$, and then we approximate $f(A^{\mS}|g_{N})$,
\beqrs
\begin{split}
f(A^{\mS}|g_{N})&\approx \int \exp\{\log{f(A^{\mS}|g_{N},\wh{\theta}) }\}p(\theta) d\theta,\\
&= \int p(\theta)\exp\{\log{f(A^{\mS})|g_{N},\wh{\theta}}\}\times \exp\{-\frac{1}{2}(\theta -\wh{\theta})^{\top}\wt{D}(\theta -\wh{\theta})\}  d\theta.
\end{split}
\eeqrs
Since $p(\theta)$ is a uniform prior probability to $\theta$, then
\beqrs
f(A^{\mS}|g_{N})\approx c_{0}f(A^{\mS}|g_{N},\wh{\theta}) \times \int \exp\{-\frac{1}{2} (\theta- \wh{\theta})^{\top} \wt{D} (\theta- \wh{\theta})\}d\theta,
\eeqrs
where $c_0$ is a constant. Considering the matrix $\wt{D}$ is symmetric, we can perform an eigenvalue decomposition on it as $\wt{D}= S^{\top} \Lambda S$, and denote the $k'$-th diagonal element of $\Lambda$ by $\lambda_{k'}$, for $k'=1,\cdots, K(K+1)/2$. Furthermore, we let a substitution $(\theta- \wh{\theta})= S^{\top}\eta$. Then, the Jacobian matrix $J(\eta)= \partial \theta/ \partial \eta= S^{\top}.$ Thus, ${\rm det}( J(\eta))=1,$ where ${\rm det}(\cdot)$ denotes the determinant function of the matrix. Furthermore,
\beqrs
\begin{split}
f(A^{\mS}|g_{N})&\approx c_{0}f(A^{\mS}|g_{N},\wh{\theta}) \times \int \exp\{-\frac{1}{2}\eta^{\top} \Lambda \eta \} {\rm det}(J(\eta)) d\eta\\
& =  c_{0}f(A^{\mS}|g_{N}, \wh{\theta}) \times \int \exp\Big\{ - \frac{1}{2} \sum_{k'=1}^{K(K+1)/2} \lambda_{k'} \eta_{k'}^{2}\Big\}d\eta\\
&=c_{0}f(A^{\mS}|g_{N}, \wh{\theta}) \times \prod_{k'=1}^{\frac{K(K+1)}{2}}\sqrt{ \frac{2\pi}{\lambda_{k'}}}\\
&= c_{0}f(A^{\mS}|g_{N}, \wh{\theta}) \times \frac{ (2\pi)^{K(K+1)/4} }{ \{ {\rm det}(\wt{D})\}^{1/2}}.
\end{split}
\eeqrs
As a result,
\beq\label{eq: fkg}
\log\{ f(A^{\mS}|g_{N})\}= \log\{f(A^{\mS}|g_{N},\wh{\theta}) \} - \frac{ 1}{2}\log{\{ {\rm det} (\wt{D})\} } +O(1).
\eeq

{\sc Step 2.} To obtain the approximation of the likelihood function, we further study the determinant of $\wt{D}$. Recall that the number of independent observations in $A^{\mS}$ is $M$. Let $\{y_r\}_{r=1}^{M}$ denote these independent observations. Then,
$\log{f(A^{\mS}| g_{N},\theta)}= \sum_{r=1}^{M} \log{f(y_r|g_{N},\theta)}.$
Note that
\beqrs
\begin{split}
\wt{D}_{k'l'}&= -\frac{ \partial^{2} \log{f (A^{\mS}|g_{N},\theta)}}{\partial \theta_{k'}\theta_{l'}}\Bigg|_{\theta= \wh{\theta}}=- \frac{ \partial^{2} \log\{ \prod_{r=1}^{M} f(y_r|g_{N}, \theta) \}}{ \partial \theta_{k'} \partial \theta_{l'}}\Bigg|_{\theta=\wh{\theta}}\\
&= -\frac{ \partial^{2}\{ 1/M\sum_{r=1}^M M\log{f(y_r|g_{N}, \theta)}\} }{\partial \theta_{k'}\theta_{l'}}\Bigg|_{\theta= \wh{\theta}}.
\end{split}
\eeqrs
As $M$ grows large, we use the weak law of large numbers on random variables, $x_{r}= M \log{f(y_r|g_{N},\theta)}, \ r= 1, \cdots, M.$
We obtain $$1/M \sum_{r=1}^{M} M\log{f(y_r| g_{N}, \theta)} \to E\left\{M \log{f(y_r| g_{N}, \theta}) \right\},$$ with high probability.
Therefore, every element in the observed Fisher information matrix is
\beqrs
\wt{D}_{k'l'}&= -\frac{ \partial^{2} E\{M \log{f(y_r|g_{N}, \theta)}\}}{\partial \theta_{k'}\partial \theta_{l'}}\Bigg|_{\theta= \wh{\theta}}= - M \frac{\partial^{2}E\{\log{f(y_r|g_{N}, \theta)} \}}{\partial \theta_{k'}\partial \theta_{l'}}\Bigg|_{\theta= \wh{\theta}}= M\wt{I}_{k'l'}
\eeqrs
where $\wt{I}_{k'l'}$ is the $(k',l')$-entry of the Fisher matrix $\wt{I}_{\theta}$ for a single observed $y_r$ ($1\le r \le M$). Thus,
\beq \label{eq: fisher}
{\rm det}( \wt{D})= (M)^{K(K+1)/2} {\rm det}(\wt{I}_{\theta}).
\eeq
To this end, according to \eqref{eq: fkg} and \eqref{eq: fisher}, we obtain
\beqrs
\log\{ f(A^{\mS}|g_{N})\}=  \log\{f(A^{\mS}|g_{N},\wh{\theta}) \}- \frac{ K(K+1)}{4}\log{M} +O(1).
\eeqrs
This accomplishes the proof.

\subsection{Proof of Proposition \ref{pro: computational}} \label{appb: computational}

To demonstrate the effectiveness of our SM-BIC, we first prove the statement regarding the computational complexity of the SM-BIC in Proposition \ref{pro: computational}.
Since the DCSBM is a generalization of the SBM, we discuss the computational complexity of the SM-BIC for DCSBM. According to the SM-BIC, there are two main procedures for determining the number of communities, including node-pair subsampling and the model selection algorithm. Therefore, we analyze the computational complexity of each procedure in detail.

First, the node-pair subsampling procedure includes two steps, where the time complexity of collecting the node set $\mS$ is $O(N)$ according to \cite{vitter1985random}, and that of forming an $N\times n$ subsampled adjacency matrix is no more than $O(Nn)$. In this way, the computational complexity of the network subsampling procedure is $O(Nn)$.

Second, perform the model selection algorithm to identify $K_0$ for the DCSBM. For each candidate $K$, the SM-BIC evaluates $K$ by the following steps.
\begin{itemize}
\item [(1)] Perform spectral clustering to the subsampled adjacency matrix $A^{\mS}$ using a truncated SVD, which takes $O(Nn)$ time complexity \citep{feng2018faster, martin2018fast}.
\item[(2)] Compute the plug-in estimator of $B$, which requires $O(Nn)$ computational complexity.
\item[(3)] Obtain the plug-in estimator of $\psi$, which has a computational cost of $O(Nn)$.
\item[(4)] Calculate the SM-BIC of $K$ with $O(Nn)$.
\end{itemize}
After repeating steps (1)--(4) $K_{\rm max}$ times, we obtain the optimal choice of the number of communities.

Since $K_{\rm max}$ is a constant, the time complexity of the SM-BIC is $O(Nn)$. Therefore, we have proved Proposition \ref{pro: computational}.

\subsection{Proof of Proposition \ref{pro: size}}\label{appb: size}

In this section, we accomplish the proof of Proposition \ref{pro: size} by the following two steps. Under the assumptions in Proposition \ref{pro: size}, we first prove that the selected node set $\mS$ covers $K_0$ blocks completely with high probability. Then, we demonstrate that the expected average degree of the subnetwork could be $\mE(d)=\Omega(\log{N})$ with high probability.

{\sc Step 1.} We first represent the event $\mS \in \mM_{K_0}$ using some simple events. Specifically, we describe the event $e=\{\mS: \ \forall \ k \in [K_0], \ \exists \ i\in \mS, g^{*}_{N,i}=k\}$ using several simple events to simply calculate its probability. Denote $e_k = \{ \mS: \sum_{i \in \mS} \bI(g^{*}_{N,i}=k)>0\}$, for $k=1,\cdots, K_{0}.$ Then, we have $e= \bigcap_{k=1}^{K_0}e_{k}.$

Then, we focus on calculating the probability of event $e$. Let $e^{c}$ denote the complement set of $e$. Then, following De Morgan's laws, $e^{c}= \bigcup_{k=1}^{K_0}e_k^{c}$. Therefore, by the property of probability measure,
\beq\label{eq: probability}
P(e^{c})\le \sum_{k=1}^{K_0} P(e_k^{c}).
\eeq
Considering random simple sampling with replacement, the probability of choosing a node from the $k$-th block is $N_{k,g^{*}_{N}}/N$ in each sampling. Then, $P(e_{k}^{c})= ( 1- N_{k,g^{*}_{N}}/N)^{n}$, for $k=1,\cdots, K_0$. As a result, according to \eqref{eq: probability},
$P(e^{c}) \le \sum_{k=1}^{K_0}P(e_k^{c}) \le  K_0 ( 1- N_{{\rm min},g_{N}^{*}}/N)^{n}$. That is, $P(e) > 1- K_0 (1- N_{{\rm min},g_{N}^{*}}/N)^{n}$, where $N_{{\rm min},g_{N}^{*}} = \min_{k}(N_{k,g_{N}^{*}})$.

Consider the subsample size $n$ such that $\epsilon\ge K_0 (1- N_{{\rm min},g_{N}^{*}}/N)^{n},$ and then, $n\ge \log(K_0/\epsilon)/\log\{(1- N_{\rm min}/N)^{-1}\}.$ Under Assumption \ref{ass: balance}, we can find a constant $c_0$ such that $N_{{\rm min},g_{N}^{*}}/N> c_0/K_0$. As a result, the subsample size $n \ge \log(K_0/\epsilon)/\log\{K_0/(K_0-c_0)\}$. If $K_0$ can go to infinity with $N$, we have $n= \Omega[\log(K_0/\epsilon)/\log\{K_0/(K_0-c_0)\}]$. This condition can be simplified by taking $\epsilon =1/N$ and $K_0=O(1)$; thus, $n=\Omega(\log{N})$ in this case. Therefore, according to the assumptions in Proposition \ref{pro: size}, we have $\mS \in \mM_{K_0}$ with high probability.

{\sc Step 2.} Consider that the network density is $\rho_N$ and under Assumptions \ref{ass: sparse}--\ref{ass: balance}, we have $\mE(d)=\mE\big\{ \sum_{i=1}^{N}\sum_{j\in \mS}A_{ij}/N\big\}= \Omega(n\rho_N)$. Furthermore, since $n=\Omega(\log{N}/\rho_N)$, we have $\mE(d)= \Omega(\log{N})$. Hence, we proved Proposition \ref{pro: size}.

\section{Theoretical proof of SM-BIC }

Here, we first establish the consistency of the SM-BIC under the SBM. Specifically, we demonstrate the claim of Theorem \ref{the: under} in Appendix \ref{appc: under}, and further give the proof of  Theorems \ref{the: likelihood} and \ref{the: over} in Appendices \ref{appc: likelihood} and \ref{appc: over}, respectively. Then, we discuss the theoretical property of the SM-BIC under the DCSBM, i.e., Theorem \ref{the: convergence}, in Appendix \ref{appc: convergence}.

\subsection{Proof of Theorem \ref{the: under}}\label{appc: under}

Without loss of generality, we start with $K=K_0-1$. To prove Theorem \ref{the: under}, we focus on analyzing the log-likelihood ratio $L_{K_0-1,K_0}$, where
$L_{K_0-1,K_0}= \max_{g_{N}\in \mC(A^{\mS},K_0-1) }\sup_{B\in \mB_{K_0-1}}\log{f(A^{\mS}| g_{N}, B)} - \log{f(A^{\mS}| g_{N}^{*},B^{*})}.$
Specifically, we accomplish the proof by following three steps. We first analyze the node assignments obtained by $K_0-1$ in detail, and then we discuss the likelihood function of $L_{K_0-1, K_0}$. Finally, we establish the upper bound for $L_{K_0-1,K_0}$.

{\sloppy
{\sc Step 1.} We discuss the community assignments based on ${\rm SBM}_K$. In the under-fitting case, $K= K_0-1$, we define a merge mechanism. First, we give the merged label vector set. Define $e_{K_0-1}= \{g_{N}\in \mC(A^{\mS},K_0-1): g_{N}= U_{k,l}(g_{N}^{*}), 1\le k\neq l\le K_0\}.$ Therefore, the assignments in $e_{K_0-1}$ merge two blocks in $g_{N}^{*}$ into a block. By Lemma \ref{lem: expectation}, without loss of generality, assume that the maximum of $G(H_{g_{N}}, B^{*})$ is achieved at $g'_{N}=U_{K_0-1, K_0}(g_{N}^{*})$. Then, we establish the corresponding merged connectivity matrix $B'\in \mB_{K_0-1}$. Define $U_{k,l}(g_{N}^{*}, B^{*})$ to represent merging blocks $k$ and $l$ in $B^{*}$ by taking weighted averages with ${\bm p}$. Specifically, if $B'= U_{K_0-1,K_0}(g_{N}^{*}, B^{*})$, then
\beqrs
B'_{u(k)u(l)}= \left\{
\begin{aligned}
&B^{*}_{kl}, \  1\le k \le l \le K_0-2;  \\
&\frac{n_{kK_0-1,g_{N}^{*}} B^{*}_{kK_0-1} + n_{kK_0,g_{N}^{*}}B^{*}_{kK_0}}{n_{kK_0-1,g_{N}^{*}} + n_{kK_0,g_{N}^{*}} }, \begin{split}
&(1\le k \le K_0-2, \\
&K_0-1 \le l \le K_0);
\end{split}
\end{aligned}
\right.
\eeqrs
where $B'_{u(k)u(l)}=B'_{u(l)u(k)}$ for $1\le k \le l \le K_0-2$. Let $\bar{O}_{K_0-1K_0-1,g^{*}}=n_{K_0-1K_0-1,g_{N}^{*}}B^{*}_{K_0-1K_0-1}$, $\bar{O}_{K_0-1K_0,g_{N}^{*}}=n_{K_0-1K_0,g_{N}^{*}}B^{*}_{K_0-1K_0}$, $\bar{O}_{K_0K_0-1, g_{N}^{*}}=n_{K_0K_0-1, g_{N}^{*}}B^{*}_{K_0K_0-1}$, and $\bar{O}_{K_0K_0,g_{N}^{*}}=n_{K_0K_0, g_{N}^{*}}B^{*}_{K_0K_0}$. Then, for $K_0-1\le k,l \le K_0$,
$$B'_{u(k)u(l)}= \frac{\bar{O}_{K_0-1K_0-1,g_{N}^{*}}+\bar{O}_{K_0-1K_0,g_{N}^{*}}+\bar{O}_{K_0K_0-1, g_{N}^{*}}+ \bar{O}_{K_0K_0,g_{N}^{*}}}{n_{K_0-1K_0-1,g_{N}^{*}}+n_{K_0-1K_0,g_{N}^{*}}+n_{K_0K_0-1, g_{N}^{*}}+ n_{K_0K_0,g_{N}^{*}}},$$
where $1\le u(k) \le K_0-1$ and $1\le u(l) \le K_0-1$ are the new block labels of communities $k$ and $l$, respectively.
}

{\sc Step 2.} We now study the log-likelihood ratio $L_{K_0-1,K_0}$. We demonstrate the following critical equation in the first step:
\beq\label{eq: maxunder}
\max_{g_{N}\in \mC(A^{\mS},K_0-1)}\sup_{B\in \mB_{K_0-1}}\log{f(A^{\mS}| g_{N},B)}=\sup_{B\in \mB_{K_0-1}}\log{f(A^{\mS} | g'_N, B)}.
\eeq
The proof of \eqref{eq: maxunder} can be accomplished in two steps. We first prove this by considering $g_{N}$ far away from $g_{N}^{*}$ and close to $g_{N}'$ (up to permutation $\tau$). Specifically, define $\mJ^{-}_{\delta_n}= \{ g_{N}\in \mC(A^{\mS},K_0-1): G(H_{g_{N}}, B^{*}) -G( H_{g_{N}'}, B^{*}) < - \delta_n \}$, where $\delta_n \to 0$ slowly. Then, we apply some useful lemmas provided earlier to prove this in another case.

{\sc Step 2.1.} For $g_{N} \in \mJ^{-}_{\delta_n}$, we prove the equality \eqref{eq: maxunder}. By Lemma \ref{lem: concentration}, there exists a constant $c_1$ such that
\beqrs
\begin{split}
&\Big| F\Big(\frac{o_{g_{N}}}{M}, \frac{n_{g_{N}}}{M}\Big) - G( H_{g_{N}}, B^{*}) \Bigg| \\
&\le c_1 \sum_{1\le k\le l\le K_0-1} \Big| \frac{o_{kl,g_{N}} }{M}-\{H_{g_{N}}B^{*}H_{g_{N}}^{\top}\}_{kl} \Big|=O_{P}(\omega_n),
\end{split}
\eeqrs
where the inequality holds because $\gamma(\cdot)$ is Lipschitz on any interval bounded away from $0$ and $1$, and recall that $\omega_n=(\rho_NN\log{n}/M)^{1/2}$. Then, for any $g_{N} \in \mJ^{-}_{\delta_n}$, we have
\beq\label{eq: omega}
\begin{split}
&\sup_{B\in \mB_{K_0-1}}\log{f(A^{\mS}|g_{N}, B)} \\
&= \sup_{B\in \mB_{K_0-1}}\log{f(A^{\mS}|g_{N}', B)} + M \Big\{ F(o_{g_{N}}/M, n_{g_{N}}/M ) - G(H_{g_{N}},B^{*})\Big\} \\
& \ \ \ \ +M \Big\{ G(H_{g_{N}},B^{*}) - G( H_{g_{N}'}, B^{*}) \Big\} + M \Big\{G(H_{g_{N}'}, B^{*}) - F\{o_{g_{N}'}/M, n_{g_{N}'}/M\} \Big\} \\
& =  \sup_{B\in \mB_{K_0-1}}\log{f(A^{\mS}|g_{N}', B)} + O_P(M\omega_{n} -M\delta_n + M\omega_{n})
\end{split}
\eeq
Hence, we obtain
\begin{align}
&\max_{g_{N}\in \mJ^{-}_{\delta_n}}\sup_{B \in \mB_{K_0-1}} \log{f(A^{\mS}| g_{N}, B)} \nonumber \\
&\le \log\Big\{ \sum_{g_{N} \in \mJ^{-}_{\delta_n}} \sup_{B \in \mB_{K_0-1}}f(A^{\mS}| g_{N},B) \Big\} \nonumber \\
&= \log\Big[\sum_{g_{N}\in \mJ^{-}_{\delta_n}} \sup_{B\in \mB_{K_0-1}} \exp\{\log{f(A^{\mS} |g_{N}, B)}\} \Big]\nonumber \\
& \le \log\Big[ \sup_{B \in \mB_{K_0-1}} f(A^{\mS}|g_{N}', B) (K_0-1)^{n}\exp\{O_{P}(2M\omega_n -M\delta_n)\}\Big] \label{eq: sumover}\\
& \le \log{\sup_{B\in \mB_{K_0-1}}f(A^{\mS}| g_{N}',B)},\label{eq: beta}
\end{align}
where \eqref{eq: sumover} is derived from \eqref{eq: omega}, and if $\delta_n \to 0$ slowly enough such that $\delta_n/ \omega_{n} \to \infty,$ we have \eqref{eq: beta}.

{\sc Step 2.2.} For $g_{N} \notin \mJ^{-}_{\delta_n},$ $|G(H_{g_{N}}, B^{*}) - G(H_{g_{N}'} , B^{*})| \to 0$. Let $\bar{g}_{N} :=\min_{\tau}| \tau(g_{N})- g_{N}'|$. Since the maximum is unique up to $\tau$, $\|H_{\bar{g}_N} - H_{g_{N}'}\|_{\infty} \to 0$. By Lemma \ref{lem: concentration},
\beqrs
\begin{split}
&P\left( \max_{g_{N}\in \tau(g_{N}')}\|W_{\bar{g}_{N}}- W_{g_{N}'}\|_{\infty} > \epsilon |\bar{g}_{N}- g'_{N}|/N\right)\\
&\le \sum_{m=1}^{N} P\left(\max_{g_{N}: g_{N}=\bar{g}_{N},|\bar{g}_{N}-g'_{N}|=m}\|W_{\bar{g}_{N}}- W_{g'_{N}}\|_{\infty} > \frac{\epsilon m}{N} \right)\\
& \le \sum_{m=1}^{N} \left\{ 2(K_0-1)^{K_0-1}N^{m}(K_0-1)^{m+2}\exp{(-c_1\rho_N^{-1}Nm)}\right\} \to 0.
\end{split}
\eeqrs
It follows for $|\bar{g}_{N}-g'_{N}| = m$, $g_{N}\notin \mJ^{-}_{\delta_n},$
\beqrs
\begin{split}
\Big\|\frac{o_{\bar{g}_N}}{M} - \frac{o_{g_{N}'}}{M}\Big\|_{\infty} &= o_P(1) \frac{|\bar{g}_N- g_{N}'|}{N} + \Big\|H_{\bar{g}_N}B^{*}H_{\bar{g}_N}^{\top} - H_{g'_{N}}B^{*} H_{g'_{N}}^{\top}\Big\|_{\infty} \\
&\ge \frac{m}{N}(c_1+ o_P(\rho_N)).
\end{split}
\eeqrs
{\sloppy Observe that $\Big\|\frac{o_{g_{N}'}}{M}- H_{g_{N}'}B^{*}H_{g_{N}'}^{\top}\Big\|_{\infty}= o_P(\rho_N)$. By Lemma \ref{lem: concentration},$ \Big\|n_{g_{N}'}/M - H_{g_{N}'}{\one}{\one}^{\top}H_{g_{N}'}^{\top}\Big\|_{\infty}=o_P(\rho_N)$. Note that $F(\cdot, \cdot)$ has a continuous derivative in the neighborhood for $(o_{g_{N}'}/M, n_{g_{N}'}/M)$. By Lemma \ref{lem: expectation},
\beqrs
\frac{\partial F\Big\{ (1-\epsilon)o_{g_{N}'}/M+ \epsilon Q, (1-\epsilon) n_{g_{N}'}/M+\epsilon q\Big\} }{\partial \epsilon} \Bigg|_{\epsilon= 0^{+}} < - c_1\rho_N <0,
\eeqrs
for $(Q, q)$ in the neighborhood of $(o_{g_{N}'}/M, n_{g_{N}'}/M)$. Hence, $F\left( o_{\bar{g}_N}/ M , n_{\bar{g}_N}/M\right) - F\left( o_{g_{N}'}/M, n_{g_{N}'}/M\right) \le - c_1 \rho_Nm/N.$
Furthermore, we obtain
\beq\label{eq: barg}
\begin{split}
&\sup_{B\in \mB_{K_0-1}} \log{f(A^{\mS}|\bar{g}_N, B)} - \sup_{B\in \mB_{K_0-1}}\log{f(A^{\mS}|g_{N}',B)}\\
&= M\Big\{ F\left( \frac{o_{\bar{g}_N}}{M}, \frac{n_{\bar{g}_N}}{M}\right)- F\left( \frac{o_{g_{N}'}}{M}, \frac{n_{g_{N}'}}{M}\right)\Big\} \le -c_1\frac{m\rho_NM}{N}.
\end{split}
\eeq
}
Then, we conclude as follows:
\begin{align}
&\max_{g_{N} \notin \mJ^{-}_{\delta_n}, g_{N} \notin \tau(g_{N}')} \sup_{B\in \mB_{K_0-1}} \log{f(A^{\mS}|g_{N}, B)}\nonumber \\
& \le \log\left\{\sum_{g_{N} \notin \mJ^{-}_{\delta_n}, g_{N} \notin \tau(g_{N}')} \sup_{B\in \mB_{K_0-1}} \log{f(A^{\mS}|g_{N}, B)}\right\} \nonumber \\
& \le \log\Bigg\{ \sum_{g_{N} \in \tau(g_{N}')} \sup_{B\in \mB_{K_0-1}}   f(A^{\mS}|g_{N}, B) \sum_{m=1}^{N} \frac{(K_0-1)^{m}N^{m}}{\exp(c_1m\rho_NM/N)} \Bigg\}\label{eq: third}\\
&\le \log\Bigg[ \sup_{B\in \mB_{K_0-1}}f(A^{\mS}|g_{N}', B) \sum_{g_{N}\in \tau(g_{N}')} \left\{ \sum_{m=1}^{N}\frac{(K_0-1)^{m}N^{m}}{\exp(c_1m\rho_NM/N)}\right\}  \Bigg]\nonumber\\
&= \sup_{B\in \mB_{K_0-1}}\log{f(A^{\mS}|g_{N}', B)}+ \log\Big\{ (K_0-1)^{K_0-1}\sum_{m=1}^{N} \frac{(K_0-1)^{m}N^{m}}{\exp(c_1m\rho_NM/N)} \Big\}\label{eq: fifth} \\
&\le  \sup_{B\in \mB_{K_0-1}}\log{f(A^{\mS}|g_{N}', B)}+ \log\left\{ (K_0-1)^{K_0}N^{2} \exp(-c_1\rho_NM/N ) \right\}\nonumber \\
&=  \sup_{B\in \mB_{K_0-1}}\log{f(A^{\mS}|g_{N}', B)}+ K_0\log(K_0-1) + 2\log{N}-c_1\rho_NM/N\nonumber\\
&<  \sup_{B\in \mB_{K_0-1}}\log{f(A^{\mS}|g_{N}', B)} \label{eq: notin}.
\end{align}
where \eqref{eq: third} is obtained by \eqref{eq: barg}, and the equality \eqref{eq: fifth} holds because the number of all community assignments in $\tau(g_{N}')$ is $(K_0-1)^{K_0-1}$. Additionally, the equality \eqref{eq: notin} results from $M/N=\Omega(n)=\Omega(\log{N}/\rho_N)$. Therefore, by \eqref{eq: beta} and \eqref{eq: notin}, we have accomplished the proof of  \eqref{eq: maxunder}.

{\sc Step 3.} We then use the conclusion in \eqref{eq: maxunder} to give the lower bound of $L_{K_0-1,K_0}$. We start by analyzing the bias of the maximum likelihood estimator of the connectivity matrix elements. Consider that $\sup_{B\in \mB_{K_0-1}}f(A^{\mS}|g_{N}',B)$ is uniquely maximized at
\begin{align}
&\wh{B}_{kl} = \frac{o_{kl,g_{N}'}}{n_{kl,g_{N}'}}=\frac{o_{kl,g_{N}^{*}}}{n_{kl,g_{N}^{*}}}= B_{kl}^{*} + O_P(\rho_NM^{-1/2}), \ \text{for} \ 1\le k\le l\le K_0-2,\label{eq: inbias} \\
&\wh{B}'_{kK_0-1}= \frac{ o_{kK_0-1,g_{N}^{*}} +o_{kK_0,g_{N}^{*}}}{n_{kK_0-1,g_{N}^{*}}+n_{kK_0,g_{N}^{*}}} = B'_{kK_0-1}+ O_P(\rho_NM^{-1/2}), \ \text{for} \ 1\le k\le K_0-2 \label{eq: outinbias}\\
&\wh{B}'_{K_0-1K_0-1} =\frac{ \sum_{k=K_0-1}^{K_0} \sum_{l=K_0-1}^{K_0} o_{kl,g_{N}^{*}}}{\sum_{k=K_0-1}^{K_0}\sum_{l=K_0-1}^{K_0}n_{kl,g_{N}^{*}}}= B'_{K_0-1K_0-1}+ O_P(\rho_NM^{-1/2}) \label{eq: outbias},
\end{align}
where the equalities \eqref{eq: inbias}, \eqref{eq: outinbias}, and \eqref{eq: outbias} are derived by Hoeffding's inequality \citep{hoeffding1963probability} presented in Lemma \ref{lem: hoeffding}. Hence, we have
\begin{align}
&L_{K_0-1K_0}=\sup_{B\in \mB_{K_0-1}}\log{f(A^{\mS}| g_{N}', B)}- \log{f(A^{\mS}| g_{N}^{*},B^{*})} \nonumber \\
&= \sum_{1\le k\le l \le K_0-2}\left\{o_{kl,g_{N}^{*}}\log{\left(\frac{\wh{B}_{kl}}{B^{*}_{kl}}\right)} + (n_{kl,g_{N}^{*}}- o_{kl,g_{N}^{*}})\log{\left(\frac{1-\wh{B}_{kl}}{1-B^{*}_{kl}}\right)}\right\}\nonumber\\
& + \sum_{k,l\in \mI} \left\{o_{kl,g_{N}^{*}} \log{\left(\frac{\wh{B}'_{u(k)u(l)}}{ B^{*}_{kl}}\right)}+ (n_{kl,g_{N}^{*}} - o_{kl,g_{N}^{*}})\log{\left(\frac{1-\wh{B}'_{u(k)u(l)}}{1-B^{*}_{kl}}\right)} \right\} \nonumber
\end{align}
where $\mI$ is the set of indices affected by the merge, $\mI=\{(k,l)\in [K_0]^{2}, K_0-1\le l \le K_0, k\le l\}$. For convenience, let
\begin{align}
X_1= \sum_{1\le k\le l \le K_0-2}\left\{o_{kl,g_{N}^{*}}\log{\left(\frac{\wh{B}_{kl}}{B^{*}_{kl}}\right)} + (n_{kl,g_{N}^{*}}- o_{kl,g_{N}^{*}})\log{\left(\frac{1-\wh{B}_{kl}}{1-B^{*}_{kl}}\right)}\right\},  \label{eq: unmerged}\\
X_2= \sum_{k,l\in \mI} \left\{o_{kl,g_{N}^{*}} \log{\left(\frac{\wh{B}'_{u(k)u(l)}}{ B^{*}_{kl}}\right)}+ (n_{kl,g_{N}^{*}} - o_{kl,g_{N}^{*}})\log{\left(\frac{1-\wh{B}'_{u(k)u(l)}}{1-B^{*}_{kl}}\right)} \right\}, \label{eq: merged}
\end{align}
where $X_{1}$ represents the bias within un-merged communities (i.e., $1\le k,l\le K_0-2$) and $X_{2}$ measure the bias within the merged communities (i.e., $(k,l)\in \mI$). That is $L_{K_0-1K_0}=X_1 +X_2$. Next, we discuss $X_1$ and $X_2$, accordingly.

First, by Taylor's expansion, we obtain
\begin{align}
X_1& = \sum_{1\le k\le l \le K_0-2}\left\{o_{kl,g_{N}^{*}}\log{\left(\frac{\wh{B}_{kl}}{B^{*}_{kl}}\right)} + (n_{kl,g_{N}^{*}}- o_{kl,g_{N}^{*}})\log{\left(\frac{1-\wh{B}_{kl}}{1-B^{*}_{kl}}\right)}\right\} \nonumber\\
& = \sum_{1\le k\le l \le K_0-2} \Bigg[ n_{kl,g_{N}^{*}}(B_{kl}^{*}+\Delta_{kl})\Big\{\frac{\Delta_{kl}}{B^{*}_{kl}}-\frac{\Delta^{2}_{kl}}{2 (B^{*}_{kl})^{2}}\Big\} \nonumber\\
& \ \ \ \ + n_{kl,g_{N}^{*}}( 1- B^{*}_{kl} - \Delta_{kl}) \left\{ \frac{-\Delta_{kl}}{1- B_{kl}^{*}}- \frac{ \Delta_{kl}^{2}}{2(1- B^{*}_{kl})^2}\right\} + O(n_{kl,g_{N}^{*}} \Delta_{kl}^{3})\Bigg] \label{eq: taylor}\\
& = \sum_{1\le k\le l \le K_0-2} \Bigg[ n_{kl,g_{N}^{*}} \Big(\Delta_{kl}+ \frac{\Delta_{kl}^{2}}{2B^{*}_{kl}} \Big) \nonumber\\
& \ \ \ \ + n_{kl,g_{N}^{*}} \left\{- \Delta_{kl}+ \frac{\Delta_{kl}^{2}}{2(1-B_{kl}^{*})}\right\}+ O(n_{kl,g_{N}^{*}} \Delta_{kl}^{3}) \Bigg]\nonumber\\
& = \frac{1}{2} \sum_{1\le k\le l \le K_0-2} \frac{n_{kl,g_{N}^{*}}(\wh{B}_{kl}- B^{*}_{kl})^{2}}{B^{*}_{kl}(1-B^{*}_{kl})} + O_P(\rho_N^{3}M^{-1/2}).\label{eq: convergence}
\end{align}
where $\Delta_{kl}= \wh{B}_{kl}- B^{*}_{kl}$ in equality \eqref{eq: taylor}, and \eqref{eq: convergence} results from \eqref{eq: inbias}. Hence, the upper bound of \eqref{eq: unmerged} is $O_P(\rho_N).$ Then, we focus on \eqref{eq: merged}. By Taylor expansion, we have
\begin{align}
X_2&=\sum_{k,l\in \mI} \Bigg[ o_{kl,g_{N}^{*}} \log{\Big\{ \frac{\wh{B}'_{u(k)u(l)}}{ B^{*}_{kl}} \Big\} }  + (n_{kl,g_{N}^{*}} - o_{kl,g_{N}^{*}})\log{\Big\{\frac{1-\wh{B}'_{u(k)u(l)}}{1-B^{*}_{kl}}\Big\}} \Bigg]\nonumber\\
&=\sum_{k,l\in \mI} n_{kl,g_{N}^{*}}\Bigg[  B'_{u(k)u(l)}\log{ \Big\{ \frac{B'_{u(k)u(l)}(1- B^{*}_{kl})}{ (1-B'_{u(k)u(l)})B^{*}_{kl} }\Big\}}\nonumber\\
&\ \ \ \  + \log{\Big\{ \frac{1- B'_{u(k)u(l)}}{1-B^{*}_{kl}}\Big\} }+O(\Delta'_{u(k)u(l)})\Bigg]\label{eq: second}\\
&= -\Omega_P(\rho_NM),\nonumber
\end{align}
where $\Delta'_{u(k)u(l)}= \wh{B}'_{u(k)u(l)}- B'_{u(k)u(l)}$. Hence, we have $L_{K_0-1,K_0}=X_1+X_2= -\Omega_P(\rho_NM)$. Therefore, we have accomplished the proof of Theorem \ref{the: under}.

\subsection{Proof of Theorem \ref{the: likelihood}}\label{appc: likelihood}

Based on the proof of Theorem \ref{the: under}, we prove the convergence of the penalized log-likelihood function $\ell(K_0)$ via the following two steps.

{\sc Step 1.} For $K=K_0$, according to \eqref{eq: maxunder}, we have
\beq\label{eq: unbias}
\max_{g_{N}\in\mC(A^{\mS},K_0)}\sup_{B\in \mB_{K_0}} \log{f(A^{\mS}|g_{N},B)} = \sup_{B\in \mB_{K_0}}\log{f(A^{\mS}| g_{N}^{*}, B)}.
\eeq
Hence, $L_{K_0,K_0}= \sup_{B\in \mB_{K_0}}\log{f(A^{\mS}| g_{N}^{*}, B)}-\log{f(A^{\mS}| g_{N}^{*}, B^{*})}$.

{\sc Step 2.} Consider that $\sup_{B\in \mB_{K_0}}f(A^{\mS}|g_{N}^{*},B)$ is uniquely maximized at
\beq\label{eq: bmse}
\wh{B}_{kl} =\frac{o_{kl,g_{N}^{*}}}{n_{kl,g_{N}^{*}}}= B_{kl}^{*} + O_P( \rho_N M^{-1/2}), \ \text{for} \ 1\le k\le l\le K_0.
\eeq
Then, similar to \eqref{eq: taylor}, by Taylor expansion, we have,
\beq\label{eq: diverge}
\begin{split}
&\sup_{B\in \mB_{K_0}}\log{f(A^{\mS}| g_{N}^{*}, B)}-\log{f(A^{\mS}| g_{N}^{*}, B^{*})}\\
&= \sum_{1\le k\le l \le K_0}\left\{ o_{kl,g_{N}^{*}}\log{\left\{ \frac{\wh{B}_{kl}( 1-B^{*}_{kl})  }{B^{*}_{kl}(1-\wh{B}_{kl})} \right\} } + n_{kl,g_{N}^{*}}\log{\left(\frac{1-\wh{B}_{kl}}{1-B^{*}_{kl}}\right)}+ O(n_{kl,g_{N}^{*}} \Delta_{kl}^{3})  \right\}\\
& = \frac{1}{2} \sum_{1\le k\le l \le K_0} \frac{n_{kl,g_{N}^{*}}(\wh{B}_{kl}- B^{*}_{kl})^{2}}{B^{*}_{kl}(1-B^{*}_{kl})} + O_P(\rho_N^3 M^{-1/2}).
\end{split}
\eeq
The last equality results from \eqref{eq: bmse}, that is $\Delta_{kl}= O_P(\rho_NM^{-1/2})$. Hence, $L_{K_0,K_0}=O_P(\rho_N)$, and this accomplishes the proof of Theorem \ref{the: likelihood}.

\subsection{Proof of Theorem \ref{the: over}}\label{appc: over}

Based on the proof of Theorem \ref{the: likelihood}, we define a log-likelihood ratio as
\beqrs
\begin{split}
\wt{L}_{K,K_0}= \max_{g_{N}\in \mC(A^{\mS},K)}\sup_{B\in \mB_{K}}\log{f(A^{\mS}|g_{N},B)}- \sup_{B\in \mB_{K_0}} \log{f(A^{\mS}|g_{N}^{*},B)}.
\end{split}
\eeqrs
To provide the upper bound of $L_{K,K_0}$, we start by discussing $\wt{L}_{K,K_0}$. Specifically, we establish the upper bound of $L_{K,K_0}$ by the following three steps. First, we introduce a set of community assignments that is formed by splitting the underlying node assignments into $K$ blocks. Second, we study the corresponding likelihood functions of $\wt{L}_{K,K_0}$. Third, based on the conclusion of $\wt{L}_{K,K_0}$, we use the preceding lemmas to accomplish this proof.

{\sc Step 1.} We first define the community assignment set by splitting the underlying $g_{N}^{*}$ into $K$ blocks. Intuitively, embedding a $K_0$-block model in a larger model can be achieved by appropriately splitting the labels $g_{N}^{*}$. Specifically, we define a subset $$e_{K}=\{g_{N} \in \mC(A^{\mS},K): \text{each row of} \ H_{g_{N}} \ \text{has at most one nonzero entry}\}.$$ Then, any $g_{N}\in e_{K}$ satisfies the following: every block in $g_{N}$ is a subset of an existing block in $g_{N}^{*}$. Accordingly, we define a surjective function as $h: [K] \to [K_0]$ describing the assignments in $H_{g_{N}}$. In other words, for any $k \in [K],$ $h(k)\in [K_0],$ and $\forall \ a \in [K_0]$, $h^{-1}(a) \in [K]$.

{\sc Step 2.} We then discuss the log-likelihood ratio $L_{K,K_0}$. Note that, in this case, $G(H_{g_{N}}, B^{*})$ is maximized at any $g_{N}' \in e_{K}$ with value $\sum_{1\le k \le l \le K_{0}}{\bm p}_{k}{\bm p}_{l}\gamma(B^{*}_{kl})$. Denote the optimal $G^{*}= \sum_{1\le k \le l \le K_{0}}{\bm p}_{k}{\bm p}_{l}\gamma(B^{*}_{kl})$. Let $\mJ_{\delta_n}^{+} = \{g_{N} \in \mC(A^{\mS},K): G(H_{g_{N}}, B^{*})- G^{*}< -\delta_n\},$ for $\delta_n \to 0$ slowly enough. Then, to analyze the log-likelihood ratio $\wt{L}_{K,K_0}$, we consider the likelihood $\sup_{B\in \mB_{K}}\log{f(A^{\mS}|g_{N},B)}$ under two cases, namely, the community assignment $g_{N}\in \mJ_{\delta_n}^{+}$ and $g_{N} \notin \mJ_{\delta_n}^{+}$.

{\sc Step 2.1 } We analyze $\sup_{B\in \mB_{K}}\log{f(A^{\mS}|g_{N},B)}$ by considering $g_{N} \in \mJ_{\delta_n}^{+}$. By Lemma \ref{lem: concentration}, we have
\beqrs
\Big|F\Big( \frac{o_{g_{N}}}{M} , \frac{n_{g_N}}{M}\Big) - G(H_{g_{N}}, B^{*})\Big| \le c_1 \sum_{1\le k\le l \le K} \Big|\frac{o_{kl,g_{N}}}{M} - (H_{g_{N}}B^{*}H_{g_{N}}^{\top})_{kl}\Big| = O_P(\omega_n).
\eeqrs
Therefore, for any $g_{N}' \in e_K$, we obtain
\begin{align}
&\max_{g_{N}\in \mJ_{\delta_n}^{+} } \sup_{B\in \mB_K} \log{f(A^{\mS}| g_{N},B)}\nonumber \\
&\le \log\Big\{ \sum_{g_{N} \in \mJ_{\delta_n}^{+} } \sup_{B\in \mB_{K}} f(A^{\mS}|g_{N}, B) \Big\} \nonumber \\
&= \log\Big[ \sum_{g_{N} \in \mJ_{\delta_n}^{+}} \sup_{B\in \mB_K} \exp{\Big\{\log{ f(A^{\mS}|g_{N}, B)} \Big\}} \Big] \nonumber \\
&\le \log\Big[\sup_{B\in \mB_{K}} f(A|g_{N}', B)(K-1)^{n} \exp{\{ O_P(2M\omega_n -M\delta_n)\}} \Big]\nonumber \\
&\le \log\Big\{\sup_{B\in \mB_{K}}f(A|g_{N}',B) \Big\} \nonumber \\
&= \sup_{B\in \mB_{K}} \sum_{1\le a\le b\le K_0}\sum_{(k,l)\in h^{-1}(a) \times h^{-1}(b)}\{ o_{kl,g_{N}'}\log{\left(\frac{B_{kl}}{1-B_{kl}}\right)} +n_{kl,g_{N}'}\log{(1-B_{kl})}\}.\label{eq: inset1}
\end{align}
Choosing $\delta_{n} \to 0$ slowly enough such that $\delta_{n}/\omega_n \to \infty$.

We further analyze equality \eqref{eq: inset1}. Let
\beqrs
\begin{split}
l_{ab}= &\sum_{(k,l)\in h^{-1}(a) \times h^{-1}(b)}\Big\{ o_{kl,g_{N}'}\log{B_{kl}} + (n_{kl,g_{N}'}- o_{kl,g_{N}'})\log{(1-B_{kl})}\Big\} \\
&+ \lambda' \Big(\sum_{(k,l) \in h^{-1}(a)\times h^{-1}(b)}n_{kl,g_{N}'} - n_{ab,g_{N}^{*}} \Big).
\end{split}
\eeqrs
Then, $\partial l_{ab}/\partial n_{kl,g_{N}'}= \log{(1-B_{kl})} + \lambda' = 0$. This implies that for $(k,l)\in h^{-1}(a)\times h^{-1}(b)$, $B_{kl}$'s are all equal. Let $B_{kl}= B_{ab}$. Hence,
\beqrs
\begin{split}
&\sum_{(k,l)\in h^{-1}(a) \times h^{-1}(b)}\{ o_{kl,g_{N}'}\log{B_{kl}} + (n_{kl,g_{N}'}- o_{kl,g_{N}'})\log{(1-B_{kl})}\}\\
&= o_{ab,g_{N}^{*}}\log{B_{ab}} + (n_{ab,g_{N}^{*}} - o_{ab,g_{N}^{*}})\log{(1-B_{ab})},
\end{split}
\eeqrs
where
$$o_{ab,g_{N}^{*}}= \sum_{(k,l) \in h^{-1}(a)\times h^{-1}(b)} o_{kl, g_{N}'} \ \ \text{and} \  \ n_{ab,g_{N}^{*}}=\sum_{(k,l) \in h^{-1}(a)\times h^{-1}(b)} n_{kl, g_{N}'}.$$
Therefore, based on \eqref{eq: inset1}, we have
\beq\label{eq: inset}
\begin{split}
&\max_{g_{N}\in \mJ_{\delta_n}^{+}}\sup_{B\in \mB_{K}} \log{f(A^{\mS}|g_{N},B)} \\
&\le \sup_{B\in \mB_{K}}\sum_{1\le a\le b\le K_0}\sum_{(k,l)\in h^{-1}(a) \times h^{-1}(b)}\left\{ o_{kl,g_{N}'}\log{\left(\frac{B_{kl}}{1-B_{kl}}\right)} +n_{kl,g_{N}'}\log{(1-B_{kl})}\right\} \\
&=\sup_{B\in \mB_{K_0}}\sum_{1\le a\le b\le K_0} \{o_{ab,g_{N}^{*}}\log{B_{ab}} + (n_{ab,g_{N}^{*}} - o_{ab, g_{N}^{*}}) \log{(1-B_{ab})}\} \\
& = \sup_{B\in \mB_{K_0}} \log{f(A^{\mS}|g_{N}^{*},B)}.
\end{split}
\eeq

{\sc Step 2.2} We investigate the likelihood function $\sup_{B\in \mB_{K}}\log{f(A^{\mS}|g_{N},B)}$ for $g_{N} \notin \mJ_{\delta_{n}}^{+}$. Treating $H_{g_{N}}$ as a vector, $\{H_{g_{N}}: g_{N} \in e_{K}\}$ is a subset of the union of some of the $K- K_{0}$ faces of polyhedron $P_{H_{g_N}}$. For every $g_{N}\notin e_{K}$, $g_{N}\notin \mJ_{\delta_{n}}^{+}$, let $g_{\perp}$ be such that $ H_{g_{\perp}} := \min_{H_{g_{N}'}: g_{N}'\in e_{K}}\|H_{g_{N}} -H_{g_{N}'}\|_{2}$. Then, $H_{g_{N}} - H_{g_{\perp}}$ is perpendicular to the corresponding $K-K_0$ face. This orthogonal implies that the directional derivative of $G(\cdot,B^{*})$ along the direction of $H_{g_{N}}- H_{g_{\perp}}$ is bounded away from 0. That is,
\beqrs
\frac{ \partial G\{(1-\epsilon) H_{g_{\perp}}+ \epsilon H_{g_{N}}, B^{*}\} }{\partial \epsilon }\Big|_{\epsilon=0^{+}} < -c_1\rho_N,
\eeqrs
for some universal positive constant $c_1$. Then, similar to the proof of Theorem \ref{the: under}, we obtain $\sup_{B\in \mB_{K}}\log{f(A^{\mS}|g_{N},B)}- \sup_{B\in \mB_{K}}\log{f(A^{\mS}|g_{\perp},B)} \le -c_1m\rho_NM/N,$
for $|g_{N}- g_{\perp}| =m$. Hence, we have
\begin{align}
&\max_{g_{N}\notin \mJ_{\delta_n}^{+}, g_{N}\notin e_{K}}\sup_{B\in \mB_{K}}\log{f(A^{\mS}|g_{N},B)}\nonumber \\
&\le \max_{g_{N}\in e_{K}}\sup_{B\in \mB_{K}} \log\Big\{f(A^{\mS}|g_{N},B) \times \sum_{m=1}^{N}(K-1)^{m}N^{m}\exp{(-c_1m\rho_NM/N)}\Big\} \nonumber
\end{align}
That is,
\begin{align}
&\max_{g_{N}\notin \mJ_{\delta_n}^{+}, g_{N}\notin e_{K}}\sup_{B\in \mB_{K}}\log{f(A^{\mS}|g_{N},B)}\nonumber\\
&\le \max_{g_{N}\in e_K}\sup_{B\in \mB_K}\log{f(A^{\mS}|g_{N},B)} + 2\log{N} + \log{K} -\frac{c_1\rho_NM}{N}\nonumber
\end{align}
Let $\mu_N= 2+ \log{K}/\log{N}- c_1\rho_NM/(N\log{N})$ and further by \eqref{eq: inset1}, we have
\beq\label{eq: outset}
\max_{g_{N}\notin \mJ_{\delta_n}^{+}, g_{N}\notin e_{K}}\sup_{B\in \mB_{K}}\log{f(A^{\mS}|g_{N},B)} \le \mu_N\log{N}+ \sup_{B\in \mB_{K_0}}\log{f(A^{\mS}|g_{N}^{*},B)}
\eeq
where the inequality \eqref{eq: outset} is obtained by \eqref{eq: inset}. To this end, according to \eqref{eq: inset} and \eqref{eq: outset}, we have $\wt{L}_{K,K_0} \le \mu_{N}\log{N}$.

{\sc Step 3.} Based on the assertion, $\wt{L}_{K,K_0} \le \mu_{N}\log{N}$, we now bound the divergence of $L_{K,K_0}$. According to \eqref{eq: diverge} in the proof of Theorem \ref{the: likelihood}, we have
\beq\label{eq: overfitting}
\begin{split}
L_{K,K_0}&=\max_{g_{N}\in \mC(A^{\mS},K)} \sup_{B\in \mB_{K}} \log{f(A^{\mS}|g_{N},B)} - \log{f(A^{\mS}|g_{N}^{*},B^{*})} \\
& \le \mu_{N} \log{N} + \sup_{B \in \mB_{K_0}}\log{f(A^{\mS}|g_{N}^{*},B)} - \log{f(A^{\mS}|g_{N}^{*},B^{*})}\\
&\le \mu_N \log{N}+ \frac{1}{2} \sum_{1\le k\le l \le K_0} \frac{n_{kl,g_{N}^{*}}(\wh{B}_{kl}- B^{*}_{kl})^{2}}{B^{*}_{kl}(1-B^{*}_{kl})} + O_P( \rho_N^{3}M^{-1/2}) \\
& = \mu_N \log{N}+ O_P(\rho_N).
\end{split}
\eeq
where the last inequality is according to \eqref{eq: diverge}. Hence, $L_{K,K_0}= O_{P}(\mu_N \log{N})$ where $\mu_N =O_P(1)$ for $n, N \to \infty$. Therefore, we have accomplished the proof of Theorem \ref{the: over}.

\subsection{Proof of Theorem \ref{the: convergence}}\label{appc: convergence}

Now, we prove the convergence of the log-likelihood ratio for the DCSBM by the following two steps.

{\sc Step 1.} For $K=K_0$, according to \eqref{eq: unbias}, we obtain
\beq\label{eq: correct}
\begin{split}
L_{K_{0},K_0}&= \max_{g_{N}\in \mC(A^{\mS},K_0)}\sup_{B\in \mB_{K_0}} \log{f(A^{\mS}| g_{N}, B, \psi^{*})}-\log{f(A^{\mS}| g_{N}^{*}, B^{*}, \psi^{*})} \\
&= \sup_{B\in \mB_{K_{0}}} \log{f(A^{\mS}| g_{N}^{*}, B, \psi^{*})}- \log{f(A^{\mS}| g_{N}^{*}, B^{*}, \psi^{*})}.
\end{split}
\eeq
{\sloppy Consider that $\sup_{B\in \mB_{K_0}}f(A^{\mS}|g_{N}^{*},B, \psi^{*})$ is uniquely maximized at $\wh{B}_{kl}=o_{kl,g_{N}^{*}}/n_{kl,g_{N}^{*}}(\psi^{*}),$ for $1\le k\le l\le K_0.$ By Lemma \ref{lem: hoeffding}, for any $s>0$, we have
\beqrs
\begin{split}
P( | \wh{B}_{kl} - B^{*}_{kl}|>s)&= P( |\rho_N(\rho_N^{-1}\wh{B}_{kl})-\rho_N\wt{B}^{*}_{kl}|>s)\\
& \le P( |\rho_N^{-1}\wh{B}_{kl}- \wt{B}^{*}_{kl} |> \rho_N^{-1} s)\\
& \le \exp\{-2 \rho_N^{-2}s^2 n_{kl,g_{N}^{*}}(\psi^{*})\}
\end{split}
\eeqrs
Hence, $\Delta_{kl}= \wh{B}_{kl}-B^{*}_{kl}= O_{P}\{ \rho_N n^{-1/2}_{kl,g_{N}^{*}}(\psi^{*})\}$, for $1\le k\le l\le K_0.$
}
{\sc Step 2.} Similar to \eqref{eq: diverge}, by Taylor expansion, we have
\beq\label{eq: converged}
\begin{split}
&\sup_{B\in \mB_{K_0}} \log{f(A^{\mS}| g_{N}^{*}, B, \psi^{*})}- \log{f(A^{\mS}| g_{N}^{*}, B^{*}, \psi^{*})}\\
&= \frac{1}{2}\sum_{1\le k \le l \le K_0} \Big[ \frac{ n_{kl,g_{N}^{*}}(\psi^{*})  \Delta_{kl}^{2}}{ B^{*}_{kl}} + O\{n_{kl,g_{N}^{*}}(\psi^{*})\Delta_{kl}^{3}\}\Big].
\end{split}
\eeq
Since $\Delta_{kl}= O_{P}\{ \rho_N n^{-1/2}_{kl,g_{N}^{*}}(\psi^{*})\}$, by \eqref{eq: correct} and \eqref{eq: converged}, we obtain $L_{K_0,K_0}= O_P(\rho_N)$. Therefore, we have accomplished this proof.

\bibliographystyle{asa}
\bibliography{jiayi}

\end{document}